\documentclass{aa}  
\usepackage{hyperref}
\hypersetup{colorlinks=True, citecolor={blue}, urlcolor={blue},linkcolor={black}}
\usepackage{graphicx}
\usepackage{txfonts}
\usepackage{lipsum}
\usepackage{subcaption}         
\usepackage{lscape}             
\usepackage{placeins}           
\usepackage{multirow} 
\usepackage{xcolor} 

\begin{document}

   \title{Characterizing and spectrally modeling embedded FUor eruptions in the near-infrared}



   \author{Jiaxun Li\inst{1,2}
        \and Tinggui Wang\inst{1,3}
        \and Zheyu Lin\inst{1,2}
        }

   \institute{Department of Astronomy, University of Science and Technology of China, Hefei 230026, People’s Republic of China\\
             \email{lijiaxun@mail.ustc.edu.cn}
            \and School of Astronomy and Space Sciences, University of Science and Technology of China, Hefei 230026, People’s Republic of China
            \and College of Physics, Guizhou University, Guiyang 550025, People’s Republic of China
            }

   \date{Received September 30, 20XX}

 
  \abstract
   {Episodic accretion in young stellar objects (YSOs) is thought to play a critical role in addressing the "luminosity problem" associated with star formation. However, optical surveys tend to bias against sources that are heavily obscured. Infrared time-domain surveys, such as unTimely WISE, facilitate the identification of such sources within the dense star formation regions of our Galaxy.}
   {
   We aim to systematically identify and characterize FUor outbursts in infrared-selected YSOs using high-resolution spectroscopy and detailed disk modeling.
   } 
   {
  We conducted follow-up high-resolution spectroscopy with Gemini South/IGRINS for four FUor candidates discovered in infrared time-domain surveys. Using a combination of photometric and spectroscopic observations, we constructed spectral energy distributions and fit them with a disk model that incorporates an actively accreting inner disk together with a passively irradiated outer disk. 
   }
   {
   All objects show CO and H$_2$O absorption bands at 2.3$\mu$m, and their positions in the Na + Ca versus CO equivalent width diagram further corroborate their classification as FUors. The best-fitting model spectra closely match both the observed spectral features and the overall continuum, providing additional confirmation of the FUor classification. The best-fit models reveal high extinction values ($A_V$ = 10–20 mag), with $M_*\dot{M}$ comparable to those of classical FUors such as FU Orionis. Among 18 sources initially selected via infrared light curves, $6-$7 out of 8 with available spectra exhibit FUor characteristics, implying a high selection efficiency.}
   {}


   \maketitle
   \nolinenumbers

\section{Introduction} \label{sec:Introduction}
The formation of main-sequence stars from young stellar objects (YSOs) is accompanied by gas accretion, often occurring in episodic outbursts \citep{1996ARA&A..34..207H,2014ApJ...795...61B,2016ARA&A..54..135H}. During such events, the optical brightness of YSOs typically increases by 5--6 magnitudes in the $B/V$ bands, corresponding to an increase of about 2 orders of magnitude in bolometric luminosity, over a span of several
months, with elevated brightness persisting for months, years, or even decades \citep{2023ASPC..534..355F}.
These episodic accretion sources are generally classified into two categories: FUors and EXors. FUors typically exhibit long-duration outbursts lasting from years to decades, whereas EXors display shorter, recurrent events on timescales of months to a few years. The two categories also present distinct spectroscopic features. FUors exhibit absorption-dominated spectra, including prominent features from CO, water vapor, and various metal lines \citep{2012ApJ...748L...5R,2018ApJ...869..146H,2024MNRAS.529L.115G}. In contrast, EXors are characterized by strong emission lines, including hydrogen and helium recombination lines, metallic collisional excitation lines, and CO emission \citep{2014prpl.conf..387A}. A comprehensive observational overview of the spectroscopic characteristics of FUors is provided by \citet{2018ApJ...861..145C}.

A long-standing issue in YSO studies is the discrepancy between the accretion rates inferred from luminosity measurements and those estimated from star-formation duration, commonly known as the luminosity problem \citep{2009ApJS..181..321E}. Episodic accretion has been proposed as a viable solution: during quiescent phases, YSOs exhibit lower accretion rates and, consequently, lower luminosities; however, during episodic outbursts, they may accrete substantial amounts of gas in short periods \citep{2006ApJ...650..956V,2010ApJ...710..470D}. In FUor-type outbursts, luminosity is thought to be primarily powered by the accretion disk.

The radiation from YSOs mainly consists of active radiative components generated by stellar activity and accretion, and passive emission radiative components by the surrounding dust and gas after absorbing this energy.
The active radiative components of YSOs arise from three main sources \citep{2006ApJS..167..256R,2007ApJ...669..483Z,2017A&A...600A..11R,2022ApJ...927..144R}: (1) stellar contraction and nuclear processes such as deuterium burning, (2) viscous dissipation in the accretion disk, and (3) magnetospheric accretion columns. In contrast, passive radiative components originate from a dusty outer disk heated by stellar irradiation. Both active and passive radiation can be reprocessed and redistributed by the surrounding envelope \citep{1976ApJ...210..377U}.

In the early days, the discovery and follow-up observations of YSO outbursts were mainly carried out in the optical band \citep{2009ApJS..181..321E,2013ApJ...768...93F}. Nonetheless, large dust extinction significantly influences the sample's completeness, particularly impacting those deeply embedded YSOs \citep{2020MNRAS.499.1805L,2023ApJ...957....8W}. In recent years, with the development of infrared time-domain surveys, many projects have also begun to use infrared bands to search and study these sources \citep{2011ApJ...733...50M,2017MNRAS.465.3011C,2021ApJ...920..132P,2024MNRAS.528.1789L,2024MNRAS.532.2683L}. \cite{2024MNRAS.532.2683L} used unTimely WISE data to search for variable sources in the sky area $295^\circ<l<350^\circ$ and $-1^\circ<b<1^\circ$, and identified numerous young stellar outburst sources by examining their color, the amplitude of their infrared light variations, and the pattern of this light fluctuation. In this study, we performed near-infrared spectroscopic follow-up observations on several FUor candidates identified by \cite{2024MNRAS.532.2683L}, specifically those labeled as V10, V191, V450, and V978, using the IGRINS instrument on Gemini South. Section \ref{sec:Light curves and spectra} describes the light curves and spectroscopic observations. We apply a disk model to near-infrared spectra and mid-infrared photometry to derive the essential physical parameters of these candidates, as discussed in Section \ref{sec:Spectral fitting and results}.
The fitting results, along with the properties of the sources identified via their infrared variability, are discussed in Section~\ref{sec:Discussion}. Lastly, in Section~\ref{sec:conclusions}, we present a summary of our main findings.
\section{Light curves and spectra} \label{sec:Light curves and spectra}

\begin{figure*}
    \centering
    \includegraphics[width=1.0\linewidth]{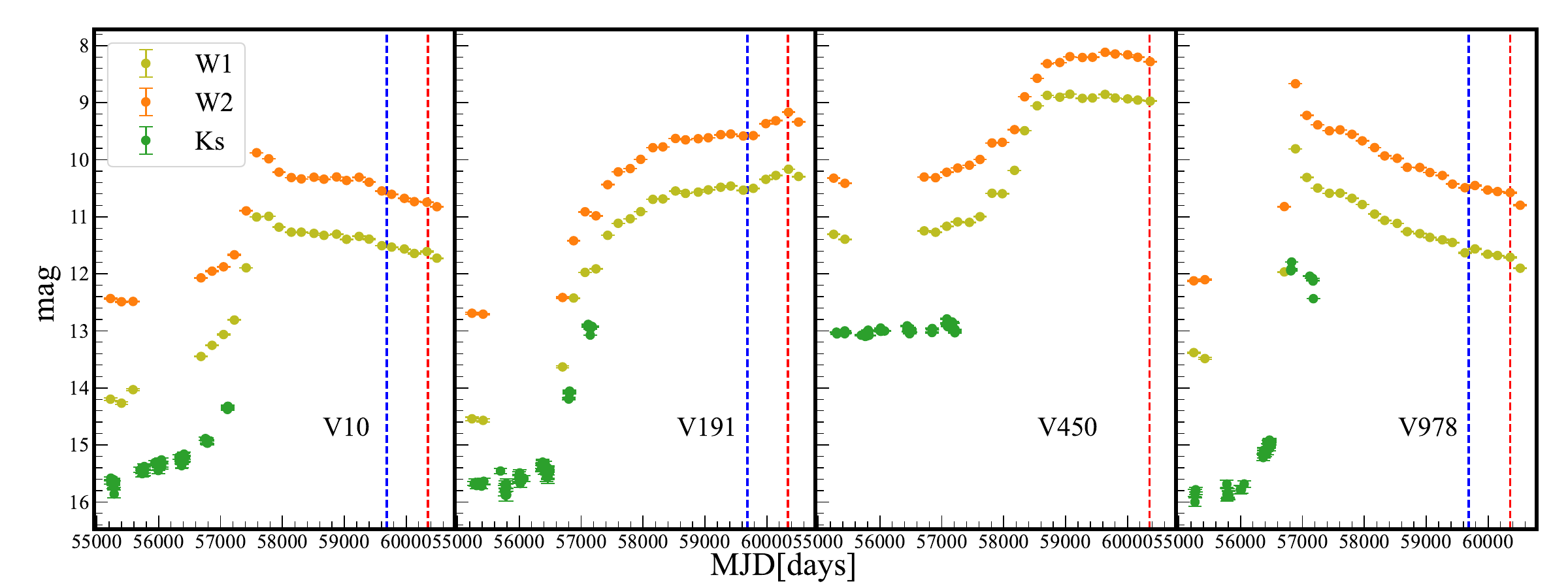}
    \caption{Near-infrared and mid-infrared light curves of the four FUor candidates. We show the lightcurves from K$_s$ (green), W1 (light green), and W2 (orange) bands\citep{2010NewA...15..433M,2014ApJ...792...30M,2019ApJS..240...30S,2023AJ....165...36M}. The vertical red and blue lines mark the dates of spectroscopic observations presented in this paper and other literature \citep{2024MNRAS.528.1769G}.}
    \label{fig:lightcurve}
\end{figure*}

\subsection{Ancillary photometry} \label{sec:Ancillary photometry}
\subsubsection{WISE} \label{sec:WISE}
The Wide-field Infrared Survey Explorer (WISE) is a NASA mission that performed an all-sky survey in the mid-infrared, repeatedly imaging the entire sky in four bands centered at 3.4, 4.6, 12, and 22 $\mu$m (W1–W4) \citep{2010AJ....140.1868W} during its first two years of operation, and subsequently in only the W1 and W2 bands following its reactivation in 2014. Its multi-epoch observing strategy makes WISE a valuable resource for time-domain photometry in the mid-infrared. In this study, we mainly utilize WISE observations to examine the infrared variability of our sources. Specifically, we use time-resolved photometric data from the NEOWISE catalogs in the W1 and W2 bands, which offer a time baseline exceeding 10 years and a typical sampling interval of about six months \citep{2011ApJ...731...53M,2014ApJ...792...30M}.

\subsubsection{VISTA} \label{sec:VISTA}
The Visible and Infrared Survey Telescope for Astronomy (VISTA) is a 4.1-m wide-field near-infrared survey telescope located at Paranal Observatory and operated by the European Southern Observatory (ESO). We use near-infrared photometric data obtained with VIRCAM, the VISTA InfraRed CAMera, which provides imaging in the Z, Y, J, H, and K$_s$ bands.

The data employed in this work are drawn from the VISTA Variables in the Vía Láctea (VVV) survey, a public near-infrared survey targeting the Galactic bulge and adjacent disk regions \citep{2010Msngr.141...24S,2010NewA...15..433M}.

For the variability analysis in this paper, we restrict the VVV data to the K$_s$ band, which has the highest observational cadence. The remaining bands do not provide adequate temporal coverage for variability studies and are therefore not considered here; measurements in those bands are reported in \cite{2024MNRAS.532.2683L}

\subsubsection{SPHEREx} \label{sec:SPHEREx}
SPHEREx, a NASA space-based all-sky spectroscopic survey mission, provides low-resolution spectra over the wavelength range of approximately 0.75--5.0 $\mu$m with a pixel scale of 6.2 arcseconds \citep{2026ApJ...999..139B}. 
We require a signal-to-noise ratio $\mathrm{flux}/\mathrm{flux\_err} > 10$ and exclude measurements with high-order SPHEREx quality flags indicating spectrophotometric contamination (flags $\gtrsim 4.3\times10^{9}$), corresponding to cases where bad pixels were involved in the spectrophotometry calculation.
An additional complication arises for the source V450. At an angular separation of 3.06 arcseconds from V450, there is a nearby contaminating object (Gaia DR3 5885045277964196096). This object has magnitudes of H = 13.181 mag and K = 12.884 mag from 2MASS \citep{2006AJ....131.1163S}, and I1 = 13.003 mag and I2 = 12.790 mag from the Spitzer GLIMPSE survey \citep{2004ApJS..154....1W}. Given the SPHEREx pixel scale is 6.2 arcseconds, the measured SPHEREx fluxes are substantially contaminated by this nearby source over several wavelength intervals. Consequently, we applied the following corrections:
(1) we discarded the spectral regions from 0.70 $\mu$m and 1.30 $\mu$m, where the contaminant dominates the SPHEREx flux;
(2) in the range 1.3–1.7 $\mu$m, we subtracted the contaminant contribution inferred from its H-band flux;
(3) we removed the 2.3–3.2 $\mu$m interval because no reliable flux measurements for the contaminant are available there;
(4) between 3.6 and 4.2 $\mu$m, we subtracted the contaminant contribution estimated from the Spitzer IRAC I1 band;
(5) between 4.2 and 5.0 $\mu$m, we subtracted the contaminant contribution estimated from the Spitzer IRAC I2 band.

The SPHEREx photometric measurements used in this work have modified Julian dates(MJD) primarily in the range 60800–60900, corresponding to approximately 500 days after the spectroscopic observations.

\subsection{Lightcurve} \label{sec:Lightcurve}
Fig.~\ref{fig:lightcurve} illustrates the light curves for the four sources. Each of these sources experienced outbursts during the NEOWISE survey, with infrared brightness rising by more than 2 magnitudes. This threshold, as indicated by \cite{2024MNRAS.532.2683L}, is used to identify YSOs exhibiting large infrared outbursts. The large amplitudes and long duration of those outbursts align with the FUor classification.

\begin{figure*}
\sidecaption
  \includegraphics[width=12cm]{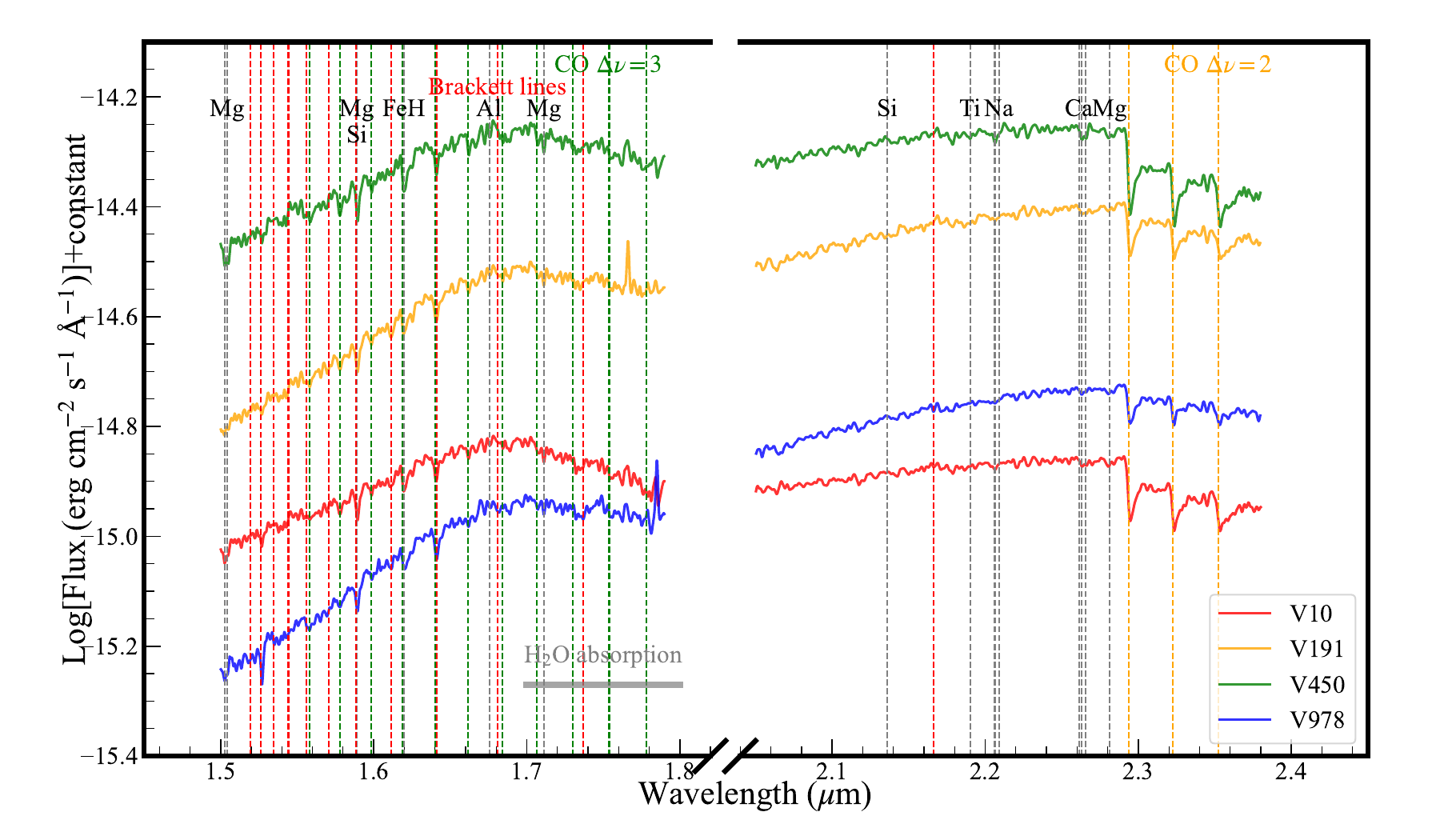}
     \caption{Near-infrared spectra of four FUor candidates.
The spectra are vertically shifted for clarity and have been convolved to a resolving power of $R = 1200$.
Prominent atomic absorption features are marked by dashed vertical lines. We show the Brackett series (red), the CO bandheads $\Delta \nu = 2$ (orange), 3 (green) and the broad H$_2$O absorption band (gray horizontal bar).}
     \label{fig:spectrum}
\end{figure*}

\subsection{Spectra}\label{sec:Spectra}
In order to confirm this classification and characterize the properties of these sources, we conducted near-infrared spectroscopic observations using the Gemini South Telescope. 
The International Gemini Observatory comprises twin 8.1-meter optical/infrared telescopes located at two of the best observing sites on Earth \citep{2018SPIE10702E..0QM}. For our observations, we used IGRINS (Immersion GRating INfrared Spectrometer, \citealt{2014SPIE.9147E..1DP}), which simultaneously covers the full wavelength range from 1.45 to 2.45~$\mu$m in a single exposure, with a resolving power of $R \sim 45{,}000$ and a slit size of $0.34^{\prime\prime} \times 0.5^{\prime\prime}$. 
The spectra were reduced using the IGRINS Pipeline Package (PLP; \citealt{kaplan2024plp}). 
They were then deblazed and corrected for velocity as described in Section~\ref{sec:Radial velocity and distance}. We performed absolute flux calibration of the spectra by
matching them to the fluxes measured by SPHEREx. For all targets except V978, the SPHEREx photometry provides a stable reference across the near-infrared bands.
For V978, however, the SPHEREx photometric points in the H and K bands exhibit unusually large scatter, preventing a reliable flux calibration. In this case, we instead calibrated the spectrum using the telluric standard star observed during the same night.
The resulting spectra are shown in Fig.~\ref{fig:spectrum}. 
Among these objects, V10, V191, and V978 were observed with X-shooter on the VLT, and their median-resolution near-infrared spectra were presented by \cite{2024MNRAS.528.1769G}; these sources correspond to L222\_1, L222\_10, and L222\_18, respectively. 
Details of the observations are listed in Table~\ref{tab:observation date}. 

All spectra exhibit the following common features, consistent with features
typically observed in FUor-type objects \citep{2018ApJ...861..145C,2025A&A...695A.130S}:
(1) Strong CO $\Delta v = 2$ absorption bands.
Prominent CO overtone ($\Delta v = 2$) absorption is detected in all spectra. This feature is widely regarded as one of the defining spectral characteristics of FUor–type sources, originating from the hot inner regions of the accretion disk.
(2) Water vapor absorption bands. 
Strong water vapor bands are present at both ends of the H-band window, shaping the H-band continuum into a characteristic triangular or “peaky” profile. Classical FUors show such a triangular H-band continuum. This feature is also considered a hallmark of FUor-type sources.
(3) Hydrogen absorption lines. 
In the H and K bands, the spectral coverage includes the hydrogen Brackett series. Clear Brackett absorption lines are observed, in particular Br$\gamma$, which is prominently detected in absorption.
(4) Weak metal absorption lines. 
Weak absorption features from neutral metals, such as Na I (2.209 $\mu$m in vacuum) and Ca I (2.261 $\mu$m), are present. These features are characteristic of late-M–type spectra and are commonly observed in FUor objects.
(5) Lack of emission lines. In contrast to EXor-type objects, which often display strong emission lines, FUor-type sources generally show little to no line emission. Consistent with this expectation, emission lines are largely absent in our spectra, although Br$\gamma$ may exhibit a P Cygni–like profile in some cases.

\begin{table}[h]
    \centering
    \caption{Dates and exposure times of Gemini South observations}
    \label{tab:observation date}
    \resizebox{\columnwidth}{!}{
    \begin{tabular}{ccccc}
        \hline
        Sources & Ra & Dec & Observation date & Exposure time \\
        & & &  & [s] \\
        \hline
        V10 & 175.7893025 & -62.35373612 & 2024-02-10 & 4800 \\
        V191 & 201.4594050 & -62.07971039 & 2024-01-31 & 2400 \\
        V450 & 237.8082806 & -53.46380964 & 2024-02-14 & 1200 \\
        V978 & 214.0749050 & -61.37314513 & 2024-02-13 & 2400 \\
        \hline
    \end{tabular}
    }
\end{table}

\section{Spectral fitting and results} \label{sec:Spectral fitting and results}

\begin{figure*}
    \centering
    \includegraphics[width=1.0\linewidth]{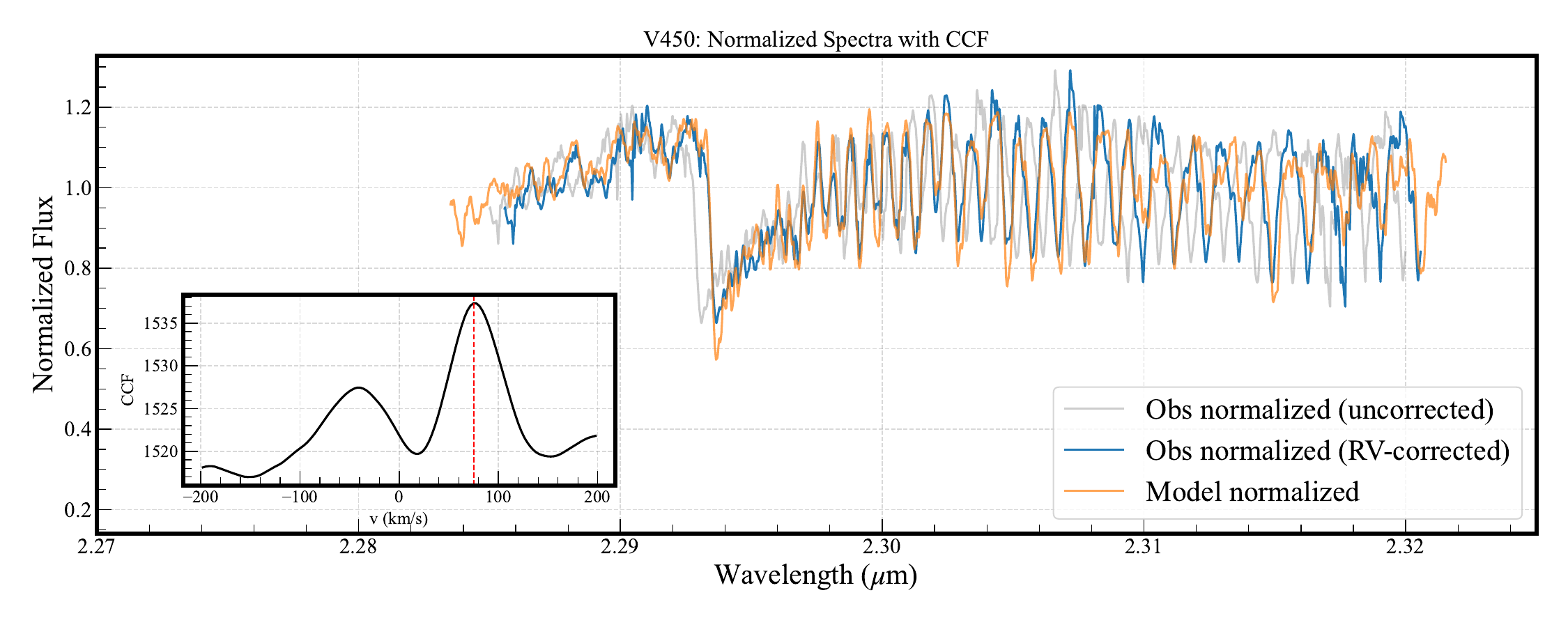}
    \caption{Example of wavelength alignment of observed and template spectra for radial velocity determination.
    The observed spectrum (gray) and the viscous model spectrum (orange) with $M_*\dot{M}=10^{-4.5} M_\odot^2/yr$, $sini\sqrt{M_{*}}=0.02$ $R_\mathrm{in}=3.5 R_\odot$, and $R_\mathrm{out}=270 R_\odot$ are continuum-normalized and shown over the wavelength range 2.2850--2.3200 $\mu$m, covering the CO $\Delta v = 2$ (2--0) absorption band. The observed spectrum after applying the radial velocity correction derived from the cross-correlation analysis is shown in blue. 
    The inset displays the CCF as a function of radial velocity, with the dashed red line indicating the velocity corresponding to the maximum correlation.
    }
    \label{fig:velocity}
\end{figure*}

\begin{figure*}
\sidecaption
  \includegraphics[width=12cm]{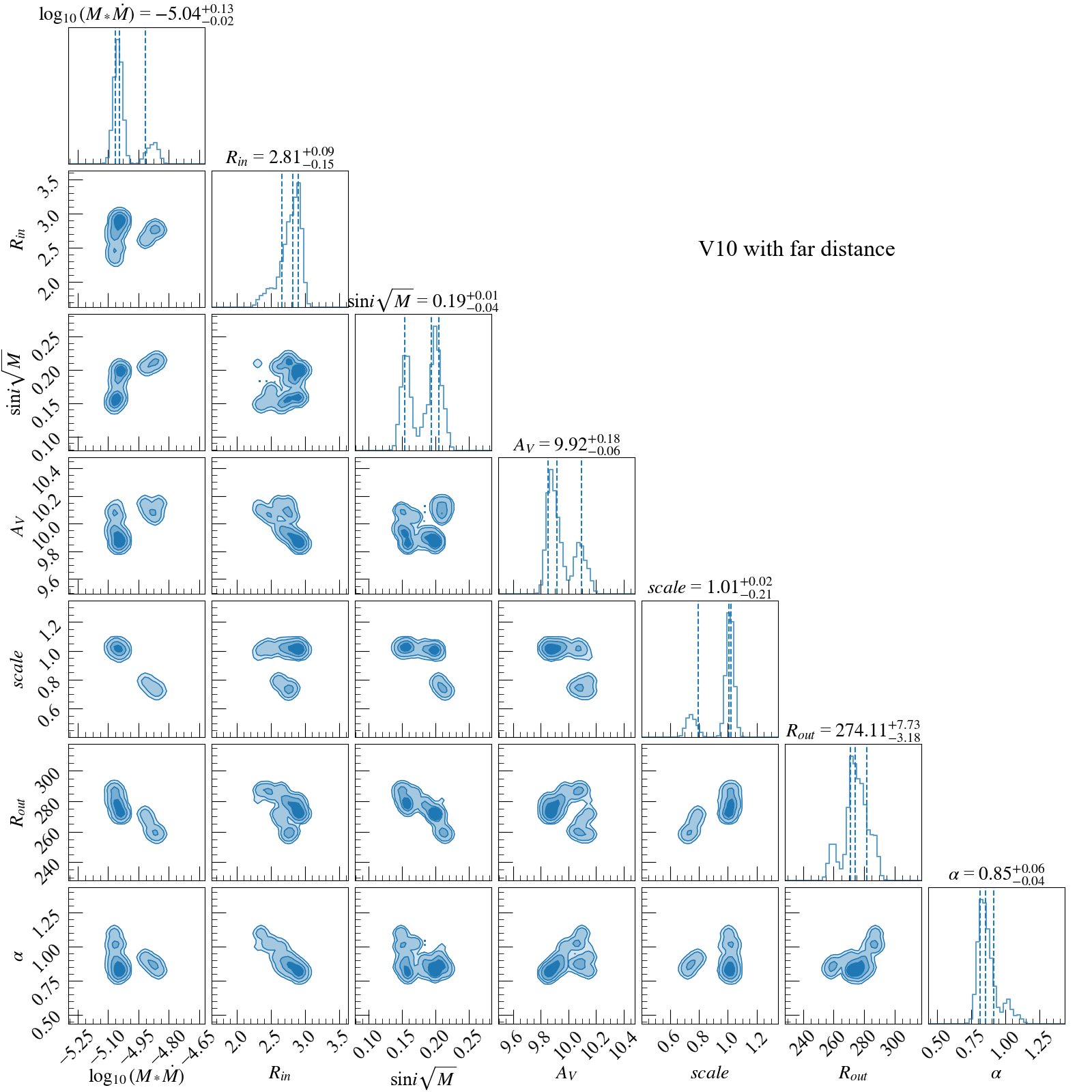}
     \caption{Parameter distributions and corner plot of V10 assuming the far distance.
The marginalized one-dimensional distributions are shown along the diagonal, where the dashed blue lines indicate the median and the 16th and 84th percentiles $\pm 1 \sigma$.
The two-dimensional panels show the joint posterior distributions, with contours enclosing 39\%, 68\%, 95\%, and 99\% credible regions.}
    \label{fig:V10_far_corner}
\end{figure*}

\subsection{Disk model} \label{Disk Model}
During episodic accretion, the near-infrared emission is dominated by the accretion disk, including both the viscous inner disk and the passively heated outer disk \citep{2022ApJ...927..144R,2022ApJ...936..152L}.
We adopt the modeling approach described in \citet{2022ApJ...936..152L} to fit the observed spectra with theoretical models and to extract the relevant physical parameters. Below we provide a brief overview of the model.
\subsubsection{Viscous accreting disk} \label{Viscous accreting disk}
We adopt the viscous accretion disk model of \cite{1973A&A....24..337S}, which assumes that the disk is in steady state, geometrically thin and optically thick. As a consequence of the differential Keplerian rotation, the disk experiences viscous heating, which leads to a temperature gradient across the radius. This distribution of temperature is expressed by
\begin{equation}
    T^4(r)=\frac{3GM_*\dot{M}}{8\pi\sigma r^3}\left[1-\left(\frac{R_\mathrm{in}}{r}\right)^{1/2} \right]
\end{equation}
where $M_*$ is the mass of the central star, $\dot{M}$ is the disk accretion rate, and $R_{\mathrm{in}}$ is the inner boundary of the disk. We denote $R_{\mathrm{out}}$ as the outer radius of the viscous disk and treat it as a free parameter. The temperature profile described by the above equation formally peaks at 
$R = 1.36\,R_{\mathrm{in}}$. 
We therefore assume that the temperature is capped at this maximum value, 
$T_{\mathrm{max}}$, and remains constant for $R < 1.36\,R_{\mathrm{in}}$, 
following \cite{2007ApJ...669..483Z} and \cite{2022ApJ...927..144R}.
To obtain the spectrum of the gas component of the viscous disk, we employ the approach outlined by \cite{2022ApJ...936..152L}. This involves using the BT–Settl grid of theoretical stellar spectra, which incorporates solar AGSS2009 abundances, to model the emission from annular regions at varying radii, each of which is associated with a particular temperature \citep{2014IAUS..299..271A}.
The atmosphere of the accretion disk should correspond to low surface gravity. In the temperature range of 1400 K to 2000 K, we utilize a surface gravity with $\log_{10} g = 3.5$, which represents the lowest gravity templates available. Between 2000 K and 24000 K, we employ templates characterized by $\log_{10} g = 1.5$. For temperatures above 24000 K or below 1400 K, we utilize a blackbody spectrum.

To account for the Doppler broadening caused by Keplerian rotation, it is essential to convolve the radiation emitted from each annulus of the accretion disk with a rotational velocity profile. The convolution kernel is specified as
\begin{equation}
    \phi(\lambda',\lambda)=\left[1- \left(\frac{\lambda'-\lambda}{\lambda_{max}} \right)^2\right]^{-1/2}
\end{equation}
\begin{equation}
    \lambda_{max}=\lambda \frac{\sin i}{c}\sqrt{\frac{GM_*}{R}}
\end{equation}
Where $i$ is the inclination angle of the disk. Assuming that the intrinsic stellar model spectrum at temperature $T$ is given by 
$F_{BT-Settl}(\lambda,T)$, the convolved flux at radius $r$ becomes
\begin{equation}
    F_{conv}(\lambda,r)=\frac{\int \phi(\lambda',\lambda)F_{BT-Settl}(\lambda',T(r))d\lambda'}{\int \phi(\lambda',\lambda)d\lambda'}
\end{equation}
The total observed spectral luminosity from the viscous accretion disk is then calculated as
\begin{equation}
    L_{vd}(\lambda)=\int_{R_{\mathrm{in}}}^{R_{\mathrm{out}}} 2\pi F_{conv}(\lambda,T(r))rdr\\
\end{equation}
\subsubsection{Passively heated disk} \label{Passively heated disk}
For the outer disk, we adopt a passively heated, flared disk geometry, in which the vertical scale height follows
\begin{equation}
\frac{H}{r} = 0.2 \left( \frac{r}{R_{\mathrm{out}}} \right)^{\beta},
\end{equation}
where $H$ is the disk scale height, $r$ is the cylindrical radius, $R_{\mathrm{out}}$ is the outer radius of the viscous disk, and $\beta$ is the flaring index \citep{2023ApJ...953...86C,2024ApJ...971...44C}.

In a passively heated disk, the temperature structure is set by radiative equilibrium between irradiation from the inner disk and re-emission by the disk surface. The incident flux at radius $r$ can be written as
\begin{equation}
F_{\mathrm{pd}}(r) \propto \frac{L_{\mathrm{irr}}}{4\pi r^{2}} \, \sin\theta ,
\end{equation}
where $L_{\mathrm{irr}}$ is the effective luminosity of the irradiation source and $\theta$ is the grazing angle of incidence.

For small angles, the grazing angle can be approximated as
\begin{equation}
\theta \simeq \frac{dH}{dr} - \frac{H}{r}.
\end{equation}
Substituting the assumed disk geometry, $H \propto r^{1+\beta}$, yields
\begin{equation}
\theta \propto \beta \, r^{\beta}.
\end{equation}
Balancing the absorbed irradiation with blackbody re-emission,
\begin{equation}
\sigma T^{4}(r) \propto F_{\mathrm{pd}}(r),
\end{equation}
the radial temperature profile of the passive disk follows
\begin{equation}
T(r) \propto r^{-\alpha},
\quad
\alpha = \frac{2 - \beta}{4}.
\end{equation}

In this framework, the temperature gradient $\alpha$ is not an independent parameter but is uniquely determined by the disk flaring index $\beta$. We treat $\alpha$ as a free parameter. We fix the outer radius of the passive disk to $1000\,R_{\odot}$. Assuming the disk radiates locally as a blackbody, the flux density emitted by the passive disk at frequency $\lambda$ can be computed by integrating the contribution from each annulus,
\begin{equation}
    L_{pd}(\lambda)=\int_{R_{\mathrm{out}}}^{1000\,R_{\odot}} 2\pi [\pi B_{\lambda}(T(r))]rdr\\
\end{equation}
where $B_{\lambda}(T)$ is the Planck function, and $R_{\mathrm{out}}$ is the inner radius of the passive disk, which is set by the outer edge of the viscous disk.
\subsubsection{Extinction from the envelope and interstellar medium} \label{Extinction from the envelope and interstellar medium}
For early-stage YSOs, the central star and its accretion disk are embedded within a collapsing, dust-filled envelope, which causes substantial extinction. In addition, because most YSOs form inside molecular clouds, the radiation they produce experiences further extinction as it passes through the surrounding large-scale material.

We adopt the extinction curve of \cite{1999PASP..111...63F} with $R_V = 3.1$ for wavelengths shorter than 3.3~$\mu\mathrm{m}$. For wavelengths longer than 3.3~$\mu\mathrm{m}$, we use the extinction prescription given in \cite{2021ApJ...916...33G}. The observed flux from the viscous accretion disk can therefore be written as
\begin{equation}
    \label{equ:13}
    F_{model}(\lambda)=\frac{(L_{vd}+L_{pd})\mathrm{cos}i}{\pi d^2_*}\times10^{-A_V\times f(\lambda)}
\end{equation}
where $d_*$ denotes the distance from the YSO to Earth, $A_V$ represents the extinction coefficient, $i$ is the inclination angle of the disk, and $f(\lambda)$ is the normalized extinction curve specified by the chosen model.

As discussed above, the model contains several unknown parameters, including the accretion rate $\dot{M}$, stellar mass $M_*$, inner radius of the viscous disk $R_{\mathrm{in}}$, outer radius of the viscous disk $R_{\mathrm{out}}$, the radial temperature gradient of the passive disk parameterized by $\alpha$, inclination angle $i$, extinction $A_V$, and distance $d_*$. 

The distance can be derived by analyzing spectral lines to obtain radial velocities, which, when combined with the Galactic rotation curve, yield an estimate of the distance to the YSO \citep{2018ApJ...856...52W}. However, this method is subject to significant uncertainties arising from both the uncertainty in the Galactic rotation curve and the peculiar velocity of the YSO relative to Galactic rotation. As shown by Equation~\ref{equ:13}, the uncertainty in distance affects the overall flux uniformly across all wavelengths, acting as a global scaling factor similar to the effect of the inclination angle through the $\cos i$ term.

In addition, the SPHEREx photometric observations were obtained approximately 500 days after the spectroscopic observations. Given the intrinsic variability of FUor-type objects, the use of non-contemporaneous SPHEREx photometry for flux calibration may introduce additional systematic offsets. We therefore absorb the combined effects of distance uncertainty, inclination, and potential temporal variability between photometric and spectroscopic observations into a single multiplicative $scale$ parameter in the fitting procedure. We therefore express the observed flux as
\begin{equation}
    F_{model}(\lambda)=\frac{(L_{vd}+L_{pd})}{\pi d^2_*}\times10^{-A_V\times f(\lambda)}\times scale.
\end{equation}
The parameters $M_*$ and $\dot{M}$ have a degenerate effect on the temperature distribution and total flux. Hence, we replace them with a single combined parameter $M_*\dot{M}$ to minimize this degeneracy. In summary, there are seven free parameters in the model: $M_*\dot{M}$, $\sin i \sqrt{M_*}$, $R_{\mathrm{in}}$, $R_{\mathrm{out}}$, $\alpha$, $A_V$, and $scale$. 

The ranges of the free parameters adopted in the fitting procedure are guided by physical considerations and previous studies of FUor-type systems. The combined parameter $M_*\dot{M}$, which has units of $M_\odot^2\,{\rm yr}^{-1}$, is allowed to vary such that $\log_{10}(M_*\dot{M})$ ranges between $-7$ and $-3$, covering the expected range of accretion rates during FUor outbursts for low-mass young stellar objects.
The inner radius of the viscous disk, $R_{\mathrm{in}}$, is constrained to $1.5$--$6\,R_\odot$, consistent with disk truncation near the stellar surface, while the outer radius $R_{\mathrm{out}}$ is limited to $60$--$300\,R_\odot$. The parameter $\sin i\sqrt{M_*}$ is restricted to $0$--$1.5\,M_\odot^{1/2}$, corresponding to typical stellar masses of low-mass YSOs. The visual extinction $A_V$ is allowed to vary between $0$ and $25$ mag, encompassing both lightly reddened and deeply embedded sources. The temperature gradient parameter $\alpha$, describing the radial temperature profile of the passive disk ($T \propto r^{-\alpha}$), is permitted to vary between $0$ and $3$, enabling the model to explore a wide range of disk flaring and irradiation geometries. Finally, the multiplicative scale parameter is allowed to vary between $0.4$ and $1.2$ to account for uncertainties in distance, inclination effects, and potential systematic offsets introduced by non-contemporaneous photometric and spectroscopic observations.

\subsection{Radial velocity and distance} \label{sec:Radial velocity and distance}

To determine the line-of-sight velocity of the spectra, we first visually inspected a set of viscous disk model spectra and selected those whose CO absorption profiles exhibit similar line spacing to the observed spectra. Both the model and observed spectra were then continuum-normalized, and a cross-correlation function (CCF) analysis was performed. The CCF was computed over the wavelength range 2.2850--2.3500 $\mu$m, which corresponds to the CO $\Delta v = 2$ (2--0) absorption band. The procedure is illustrated in Fig.~\ref{fig:velocity}.

The measured radial velocities were first corrected to the barycentric reference frame based on the time of observation, and subsequently transformed to the local standard of rest (LSR) frame. 
With the source's Galactic coordinates and the $V_\mathrm{LSR}$ derived from the observed spectra, we estimate the kinematic distance using a Galactic rotation model following \cite{1996MNRAS.283.1102P}, with updated Galactic parameters from \cite{2014ApJ...783..130R} (see \cite{2018ApJ...856...52W} for further details). 
The derived kinematic distances for each source are listed in Table~\ref{tab:parameter}. 
For several sources, both near and far kinematic distance solutions are feasible; however, V10 admits only the far distance solution.

\subsection{Derived physical parameters} \label{Derived physical parameters}

The model incorporates seven free parameters: $M_*\dot{M}$, $\sin i \sqrt{M_*}$, $R_{\mathrm{in}}$, $A_{V}$, $scale$, $R_{\mathrm{out}}$, and $\alpha$.

To derive the physical properties for each source, we fit the observed near-infrared spectra and the SPHEREx measurements with the corresponding model spectra. For a consistent comparison, we rebin both the observational and model spectra onto a common logarithmic wavelength grid, where the ratio between successive wavelength pixels is fixed at 1.00002. For the SPHEREx data, after applying the quality cuts described in Section~\ref{sec:Light curves and spectra}, we evaluate the model prediction for each SPHEREx spectral element by averaging the model flux across the wavelength interval associated with that SPHEREx bandpass. This approach ensures that the comparison between the model and the SPHEREx observations correctly incorporates the finite spectral resolution of SPHEREx.

Parameter estimation is performed using a Markov chain Monte Carlo (MCMC) approach. The details of the likelihood function, priors, and sampling strategy are described in Appendix~\ref{sec:Fitting procedure}. For all sources except V10, two possible kinematic distances (near and far) are available; therefore, we perform the fitting procedure independently for both distance solutions. In the subsequent fitting, we excluded the far distance for V450 and the near distance for V978.

To validate our fitting methodology, we apply the same procedure to the well-studied FU Orionis. We use a high-resolution spectrum observed with IGRINS at the 4.3-m Discovery Channel Telescope (DCT) at Lowell Observatory on 19 January 2019 \citep{2025PASP..137c4505S}. The spectrum is processed following the same steps described above. Since all SPHEREx data points for FU Orionis are flagged with values exceeding $4.29\times10^{9}$, indicating unreliable spectrophotometry, we do not use SPHEREx measurements for this source. Instead, the flux calibration is performed using the telluric standard star observed during the same night, following the same procedure adopted for V978. In addition, FU Orionis falls within the masked region of the kinematic distance uncertainty model (160° < $l$ < 200°), where kinematic distances are known to be highly uncertain. 
Consequently, we do not attempt a kinematic distance determination. Instead, we adopt the distance inferred from the parallax measurement reported by \cite{2021A&A...649A...1G}, as compiled in the SIMBAD database \citep{2000A&AS..143....9W}, yielding a value of 408 pc for FU Ori. This estimate is consistent with recent literature determinations, such as those reported by \cite{2018AJ....156...84K} and \cite{Roychowdhury_2024}.

\begin{figure*}
    \centering
    \includegraphics[width=1.0\linewidth]{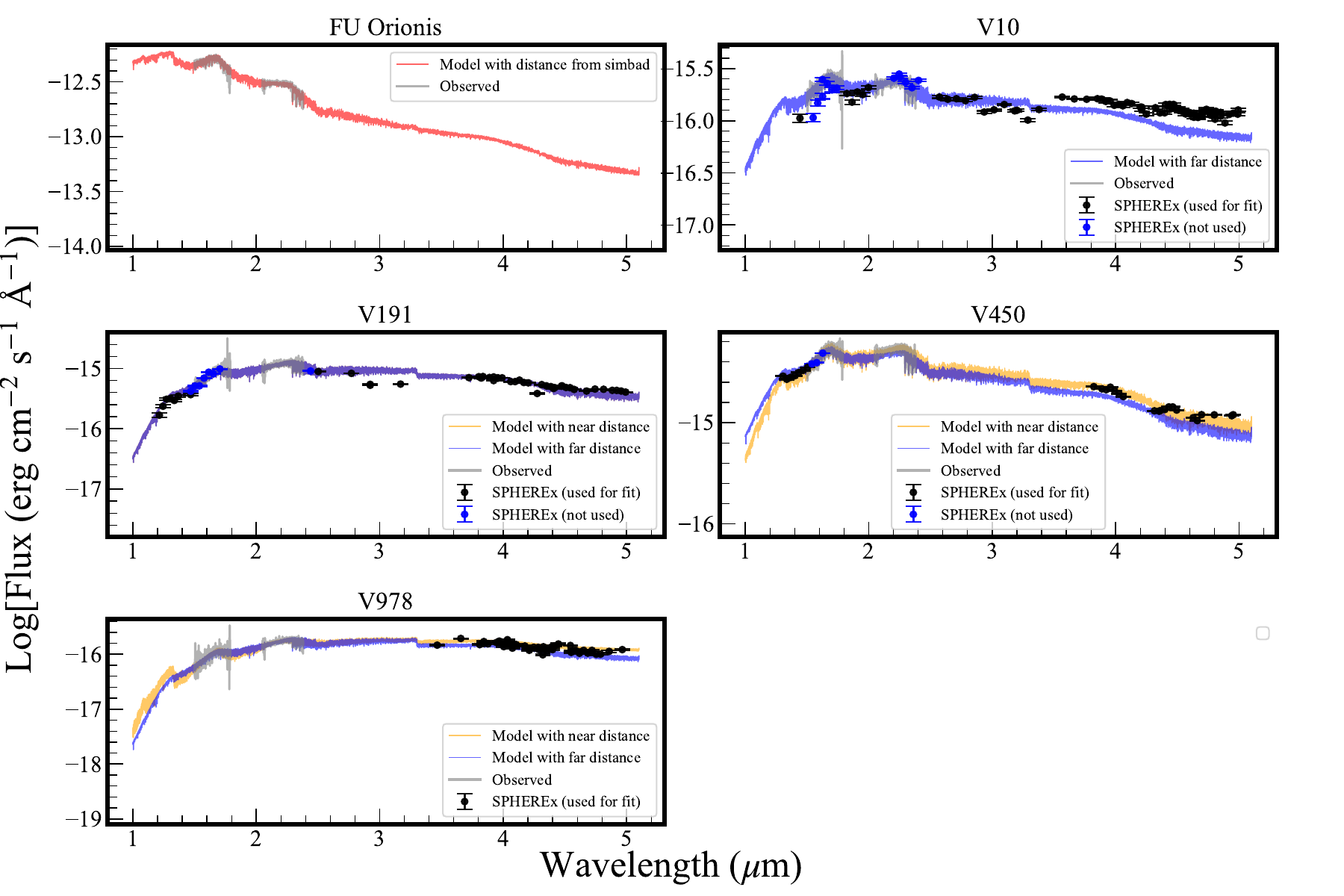}
    \caption{ 
    Comparison between the observed spectra and the corresponding best-fit model spectra for the sample objects. The gray lines display the observed near-infrared spectra in the $H$- and $K$-bands. The colored lines show the model spectra derived under different distance assumptions: the distance estimated from the parallax measurement compiled in the SIMBAD database (red), the near-distance solution (orange), and the far-distance solution (blue). Black circles with error bars mark the SPHEREx spectral data points included in the fit, whereas blue circles indicate SPHEREx measurements excluded from the fitting procedure because they are used for flux calibration. All spectra are plotted on a logarithmic flux scale.
    }
    \label{fig:sed}
\end{figure*}

For illustration, we present the posterior distributions (corner plot) for source V10 in Fig.~\ref{fig:V10_far_corner}. V10 has only a far-distance solution, while the corner plots for the remaining sources and distance solutions are provided in Appendix~\ref{sec:Distribution, corner diagrams, and spectral Fits} (Figs.~\ref{fig:corner_A1}--\ref{fig:corner_A4}). The best-fitting model spectra for all sources are shown in Fig.~\ref{fig:sed}, and zoomed-in views of selected spectral regions for the four sources and FU Orionis are shown in Fig.~\ref{fig:V10_oms} and in Appendix~\ref{sec:Distribution, corner diagrams, and spectral Fits} (Figs.~\ref{fig:V191_oms}--\ref{fig:FU Orionis_oms}). We note that for parameters with asymmetric or extended tails, the median and associated scatter should be interpreted with caution.

Overall, the models reproduce the observed spectra—both the SPHEREx photometry and our high-resolution data quite well, with good consistency in the absorption features and the general spectral shape for most objects. A notable exception is source V10, where the modeled mid-infrared flux is systematically lower than the SPHEREx measurements. This mismatch likely arises from several causes. First, the mid-infrared emission is largely produced by the passively heated outer disk, for which we assume a simplified geometric prescription, $H/r = 0.2 (r/R_{\mathrm{out}})^{\beta}$, that may not accurately represent the true disk structure or temperature profile. Second, V10 may be embedded in a surrounding envelope whose warm dust adds mid-infrared emission that is not accounted for in our disk-only model. Third, additional processes such as small-scale surface inhomogeneities in the disk or localized heating events may also contribute. Given non-contemporaneous observations and the simplified nature of our passive disk treatment, we refrain from a detailed analysis of the origin of this mid-infrared flux discrepancy in the present work. We also conclude that the far-distance solution for V450 fails to adequately reproduce the observed spectrum. As further illustrated by the associated corner plot in Appendix~\ref{sec:Distribution, corner diagrams, and spectral Fits} (Fig.~\ref{fig:corner_A2}), this solution implies $\log_{10}(M_*\dot{M}) \gtrsim -3$, which is markedly higher than the values typically derived for FUor-type objects, and is therefore rejected. In the case of V978, the near-distance solution produces model absorption features that are consistently weaker than those observed, and the corresponding accretion rate ($\log_{10}(M_*\dot{M}) \sim -6.4$ in Fig.~\ref{fig:corner_A3}) is far below what is expected for FUor outbursts; this solution is likewise discarded. As a result, V191 is the only source for which both the near- and far-distance solutions remain compatible with the observational constraints. In summary, after removing the non-physical or disfavored distance solutions, the final fitted physical parameters for all sources are reported in Table~\ref{tab:parameter}.
\begin{table*}
    \centering
    \small
    \caption{Parameter fitting results}
    \label{tab:parameter}
    \resizebox{\linewidth}{!}{
    \begin{tabular}{ccccccccccc}
        \hline
        Sources & Radial & LSR & Distance & ($\log_{10} (M_*\dot{M})/[M_\odot^2/yr]$) & $R_\mathrm{in}$ & $\sin i\sqrt{M_{*}}$ &$A_V$ &$scale$ &$R_\mathrm{out}$ & $\alpha$\\
        
        & velocity& velocity & & & & & & & &  \\
        
        & [km/s] & [km/s] & [kpc] & &[$R_\odot$] & [$\sqrt{M_\odot}$] & & & [$R_\odot$] &\\
        \hline
        V10 & 17.99 & 25.19 & 9.15 (far)  & -5.04$_{-0.02}^{+0.13}$ & 2.81$_{-0.15}^{+0.09}$ & 0.19$_{-0.04}^{+0.01}$ & 9.92$_{-0.06}^{+0.18}$ & 1.01$_{-0.21}^{+0.02}$ & 274.11$_{-3.18}^{+7.73}$ & 0.85$_{-0.04}^{+0.06}$ \\

        \multirow{2}{*}{V191} & \multirow{2}{*}{-50.96} & \multirow{2}{*}{-36.80} & 2.81 (near)  & -4.57$_{-0.01}^{+0.01}$ & 2.87$_{-0.01}^{+0.03}$ & 0.18$_{-0.00}^{+0.00}$ & 16.18$_{-0.01}^{+0.01}$ & 0.42$_{-0.01}^{+0.00}$ & 93.74$_{-1.52}^{+1.32}$ & 2.73$_{-0.28}^{+0.08}$ \\
        & & & 7.18 (far) & -3.78$_{-0.03}^{+0.02}$ & 5.39$_{-0.20}^{+0.15}$ & 0.25$_{-0.00}^{+0.00}$ & 16.17$_{-0.01}^{+0.02}$ & 0.82$_{-0.02}^{+0.03}$ & 168.28$_{-1.87}^{+1.94}$ & 2.42$_{-0.10}^{+0.18}$\\

        V450 & -74.95 & -45.66 & 3.05 (near)  & -4.25$_{-0.01}^{+0.05}$ & 5.89$_{-0.08}^{+0.02}$ & 0.31$_{-0.00}^{+0.00}$ & 11.68$_{-0.00}^{+0.01}$ & 1.00$_{-0.01}^{+0.01}$ & 86.94$_{-0.37}^{+0.22}$ & 3.00$_{-0.00}^{+0.00}$ \\

        V978 & -47.97 & -30.48 & 9.29 (far)  & -4.16$_{-0.02}^{+0.01}$ & 1.51$_{-0.01}^{+0.01}$ & 0.21$_{-0.00}^{+0.00}$ & 19.28$_{-0.01}^{+0.03}$ & 0.40$_{-0.00}^{+0.01}$ & 98.93$_{-1.17}^{+0.80}$ & 0.63$_{-0.00}^{+0.00}$ \\

        FU Orionis & 32.98 & 2.90 & 0.408 (simbad\tablefootmark{a})  & -4.87$_{-0.04}^{+0.01}$ & 1.50$_{-0.00}^{-0.00}$ & 0.55$_{-0.01}^{+0.00}$ & 2.87 $_{-0.02}^{+0.00}$ & 0.79$_{-0.01}^{+0.05}$ & 281.23$_{-21.48}^{+15.19}$ & 0.72$_{-0.69}^{+1.57}$ \\
        
        \hline
        \multicolumn{8}{l}{\textbf{Notes.}} \\
        \multicolumn{8}{l}{$^{a}$ Based on its coordinates, it is known that its distance cannot be calculated using the galactic rotation }\\
        \multicolumn{8}{l}{curve. The distance is derived from from the parallax listed in the SIMBAD database}\\
    \end{tabular}
}
\end{table*}

\begin{figure*}
    \centering
    \includegraphics[width=1.0\linewidth]{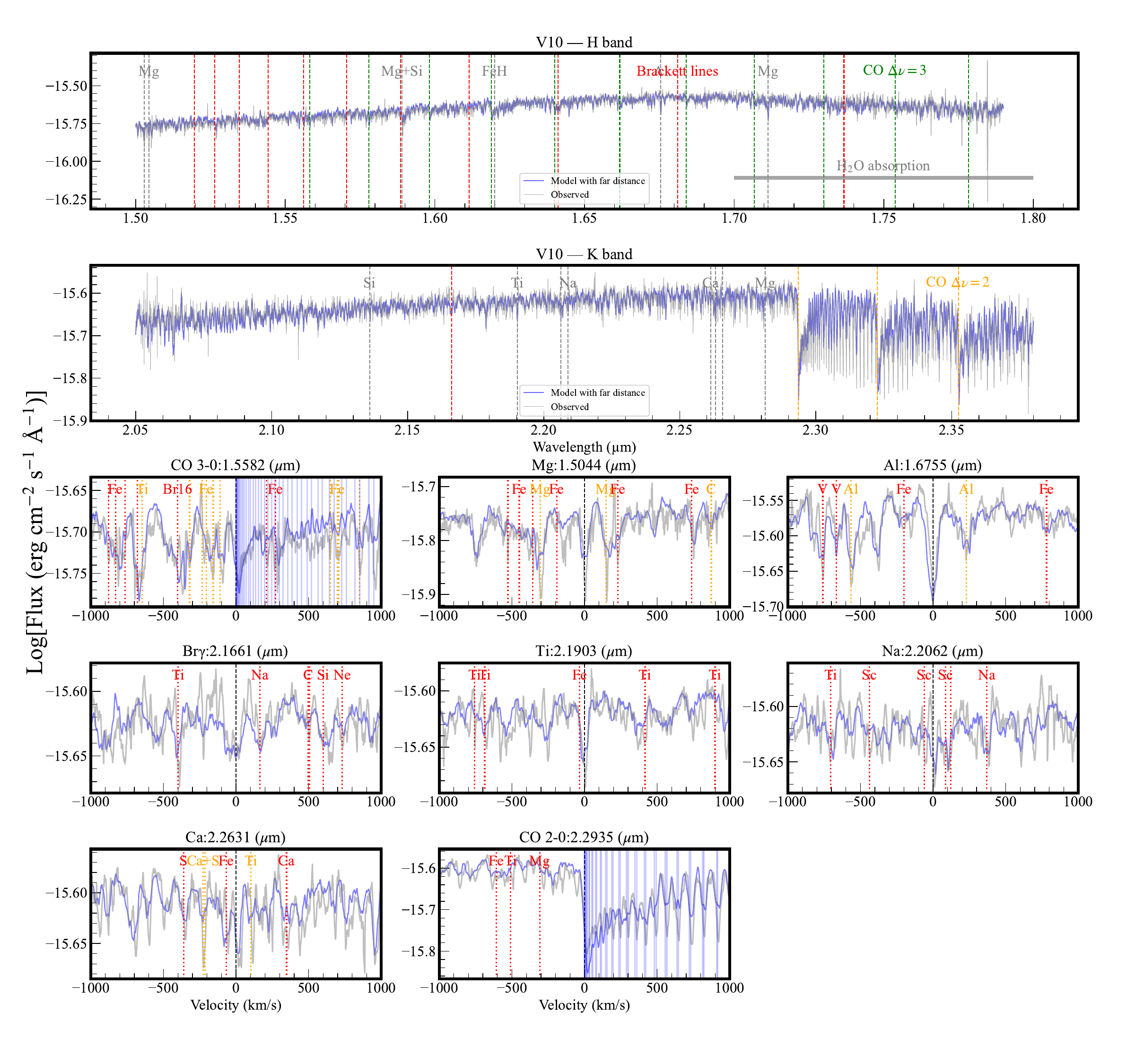}
    \caption{
    Near-infrared H- and K-band spectra of V10 compared with best-fitting disk models.
    Top panel: H band. Second: K band.
    The gray curve shows the observed spectrum after telluric and flux calibration, while colored curves represent model spectra computed using different distance assumptions: red for the distance derived from parallax measurement compiled in the SIMBAD database, orange for the near kinematic distance, and blue for the far kinematic distance (when applicable).
    Vertical dashed lines mark prominent atomic transitions (gray), Brackett lines (red), and CO overtone bandheads (green), as labeled.
    The shaded region in the H band indicates strong H$_2$O absorption.
    Lower panels: Zoomed-in views of selected spectral lines in velocity space, centered on the rest wavelength of each transition. The vertical dashed line at zero velocity marks the line center, while nearby atomic and molecular transitions are indicated by vertical markers. Atomic lines are shown in red and orange to visually separate closely spaced transitions and their labels, while molecular transitions (blue) shown in the first and eighth lower panels—correspond to different rotational transitions within the same vibrational band. Observed spectra are shown in gray and model spectra follow the same color coding as in the upper panels.
    }
    \label{fig:V10_oms}
\end{figure*}

\section{Discussion} \label{sec:Discussion}
\subsection{Validation with FU Orionis} \label{sec:Validation with FU Orionis}

FU Orionis is one of the best-studied FUor systems and therefore serves as an ideal benchmark source for validating our spectral fitting framework. By applying the same fitting procedure to its near-infrared spectrum, we are able to assess the reliability and limitations of our method in comparison with well-established results from the literature \citep{2011arXiv1106.3343H,2024ApJ...973L..40C}.

For the mass accretion rate, previous studies by \citet{2020MNRAS.495.3494Z} (see their Fig. A1) adopt a stellar mass of about $0.6\,M_\odot$ and infer an accretion rate of about $3.8\times10^{-5}\,M_\odot\,\mathrm{yr}^{-1}$, corresponding to $\log_{10}(M_*\dot{M}) \approx -4.64$. Our best-fit value, $\log_{10}( M_*\dot{M}) = -4.87$, is lower by only about $0.1$ dex. This small offset can be naturally explained by differences in flux calibration. In this work, the spectrum is flux-calibrated using the telluric standard star observed during the same night, such that the calibrated flux reflects the instantaneous brightness of FU Orionis at the time of the spectroscopic observation. Converting our calibrated spectrum to equivalent 2MASS magnitudes yields $H=5.92$ mag and $K=5.37$ mag, which are fainter than the original 2MASS measurements ($H=5.70$ mag, $K=5.16$ mag). This behavior is consistent with the long-term photometric fading observed in the ASAS-SN survey, indicating that the difference in $\log_{10}(M_*\dot{M})$ primarily reflects intrinsic variability rather than a systematic bias in our modeling.

In contrast, parameters that are more sensitive to short-wavelength constraints show larger discrepancies. Our fitted inner disk radius ($R_\mathrm{in}\approx1.5\,R_\odot$) is smaller than the $\sim3.5\,R_\odot$ reported by \cite{2020MNRAS.495.3494Z} , and the derived extinction ($A_V\approx2.87$) is higher than the commonly adopted range of $A_V\sim1.5$--2.0 \citep{2007ApJ...669..483Z,2024ApJ...973L..40C}. These differences are most likely caused by the lack of optical photometric and spectroscopic data in our fitting. Extinction is only weakly constrained in the near-infrared, and $R_\mathrm{in}$, which traces the hottest inner disk regions, is strongly coupled to the optical continuum level and line diagnostics. Without optical constraints, both parameters may therefore suffer from systematic uncertainties.

Adopting an inclination range of $37^\circ$--$55^\circ$ \citep{2020ApJ...889...59P,2007ApJ...669..483Z} and a stellar mass of $0.6\,M_\odot$, the expected range of $\sin i\sqrt{M_*}$ is 0.47--0.63, in excellent agreement with our fitted value of 0.55. In addition, the distance to FU Orionis is derived from parallax measurement compiled in the SIMBAD database rather than derived from the Galactic rotation curve, resulting in a relatively small systematic uncertainty in the scale parameter. The expected scale range of 0.57--0.80 is consistent with our best-fit value of 0.79.

FU Orionis lacks usable SPHEREx measurements, and we also examined contemporaneous NEOWISE photometry at the epoch of the spectroscopic observation. However, the NEOWISE measurements are severely inconsistent with the untimely WISE data, most likely due to saturation effects. We therefore do not include mid-infrared constraints for this source, which may lead to unreliable estimates of the outer disk radius and viscosity parameter. 

Overall, our benchmark test demonstrates that our spectral fitting approach robustly constrains key parameters such as $\log_{10}(M_*\dot{M})$ and $\sin i\sqrt{M_*}$ using near-infrared spectra alone. Parameters such as $R_\mathrm{in}$ and $A_V$ remain more susceptible to systematic uncertainties in the absence of optical data. In the case of FU Orionis specifically, $R_\mathrm{out}$ and $\alpha$ could not be reliably determined due to the lack of mid-infrared coverage, whereas for the other sources in our main sample, SPHEREx data provide sufficient constraints on these parameters.
\begin{figure}
    \centering
    \includegraphics[width=1.0\linewidth]{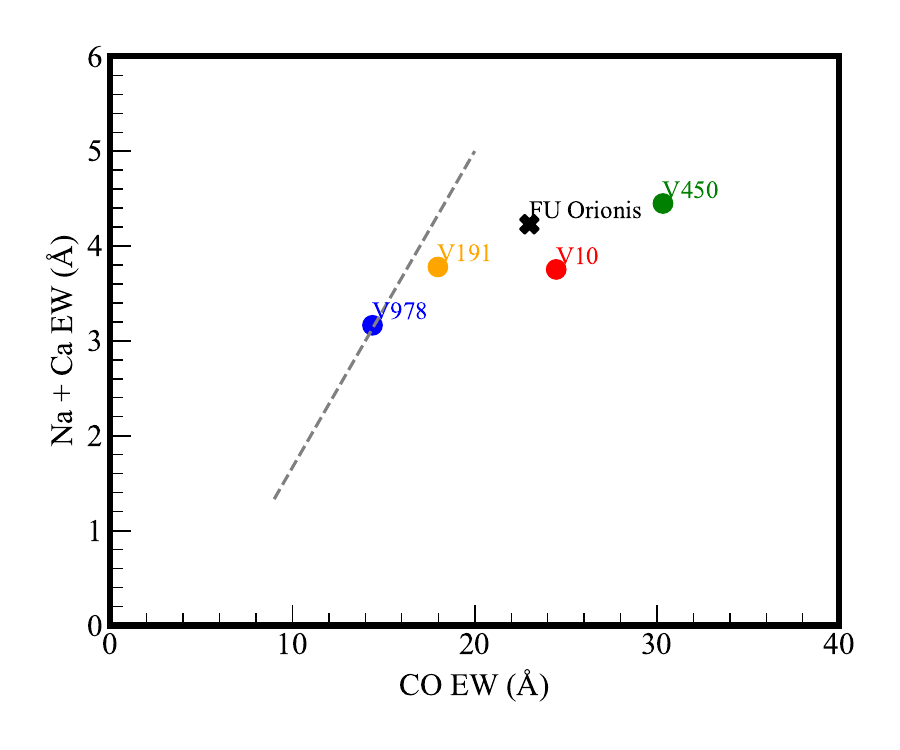}
    \caption{Equivalent-width diagram of Na I + Ca I versus CO for the FUor template and the observed candidates. The dashed line represents the empirical criterion defined by \cite{2018ApJ...861..145C} to separate FUor-type objects from other young stellar objects. Objects lying below the dashed line show enhanced CO absorption relative to Na I and Ca I and are therefore consistent with FUor-like near-infrared spectra. }
    \label{fig:ew_NaCa_vs_CO}
\end{figure}
 
\subsection{Properties and spectral characteristics of our FUor candidates} \label{Properties and spectral characteristics of our FUor candidates}

For the accretion rates, we can only constrain the combination $M\dot M$ from our spectral fitting. The four sources in our sample have $\log_{10}(M\dot M)$ values ranging from $-5.1$ to $-3.7$. Assuming a stellar mass of $0.5\,M_\odot$, the inferred accretion rates fall within the range typical for classical FUor sources, indicating that our discoveries are consistent with expected FUor accretion activity.

Regarding extinction, our sources exhibit significantly higher $A_V$ values compared to FUor sources identified via optical surveys, with even the lowest-extinction object having $A_V = 10$ mag. This highlights the importance of infrared variability searches, which enable the identification of heavily obscured FUor systems that would otherwise be missed in optical surveys. Consistent with this picture, only one source in our sample (V10) has a Gaia counterpart ($G = 20.72$ mag), while the remaining sources lack Gaia detections, underscoring the severe optical obscuration affecting these objects \citep{2023A&A...674A...1G}.

The H- and K-band spectra of our sources display characteristic FUor features, including CO $\Delta v=2$ absorption, strong H$_2$O absorption, Brackett line absorption, and weak metallic lines such as Na I (2.209 $\mu$m in vacuum) and Ca I (2.261 $\mu$m). Comparison with the multi-temperature photosphere models shows good agreement across the H--K spectral range, capturing both line strengths and subtle continuum variations. While V10 and FU Orionis exhibit weak P Cygni profiles in Br$\gamma$ (see Fig~\ref{fig:V10_oms} and Fig~\ref{fig:FU Orionis_oms}), which may indicate disk winds, confirmation requires additional J-band or optical spectroscopic data, see \cite{2024ApJ...971...44C,2023ApJ...958L..27H}, and  \cite{2025ApJ...988...77H}.

We additionally assess the FUor classification of our targets using the equivalent width (EW) diagnostic introduced by \cite{2018ApJ...861..145C}, which compares the summed EW of the Na+Ca lines to that of the CO $v=2-0$ overtone bandhead. This EW diagram has the advantage of being largely independent of inclination and distance, as the selected lines arise from similar radii and are subject to similar veiling. Adopting the procedure of \cite{2022ApJ...936..152L}, we first degraded our spectra to a resolution of 1200 prior to measuring the EWs. As illustrated in Fig.~\ref{fig:ew_NaCa_vs_CO}, all objects fall on the FUor side of the empirical boundary EW$_{\rm CO} = 3 \times {\rm EW}_{\rm Na+Ca} + 5$\,\AA, with the exception of V978, which lies close to the dividing line. We caution that our approach to determining EW$_{\rm CO}$ tends to yield slightly lower values than those reported in earlier studies (see \cite{2022ApJ...936..152L}); therefore, we regard all sources as consistent with the FUor EW criterion.

\subsection{Confirmation rate of FUor candidates selected from infrared light curves} \label{sec:Confirmation rate of FUor candidates selected from infrared light curves}

\citet{2024MNRAS.532.2683L} identified a total of 18 FUor candidates, together with one EXor candidate, based on distinctive infrared light-curve morphologies. To date, near-infrared spectra are available for 8 of these 18 FUor candidates. Among them, one source is classified as an EXor (V1298). Another object, V148, was previously characterized as an EXor (see \cite{2021MNRAS.504..830G}) but has recently been found to show a transition from CO emission to CO absorption, a hallmark of FUor-type spectra (Li et al., in prep.). In addition, V1707 is a rare object exhibiting strong aluminium monoxide (AlO) absorption while still retaining otherwise typical FUor spectral characteristics. The remaining sources—V10, V191, V450, V978, and V1349—exhibit conventional FUor spectral features.

In our analysis, we confirm FUor-type spectra for six sources including four sources in this paper. Although V148 does exhibit CO absorption in our observations, our preliminary reduction suggests that the absorption lines are significantly broader than those observed in the FUor sources analyzed in this work. If interpreted within the FUor framework, this would imply a relatively large value of $\sin i \sqrt{M_*}$. Moreover, the infrared variability amplitude of V148 appears weaker compared to the confirmed FUor sources in our sample. Given these ambiguities, and in light of newly approved K-band spectroscopic observations, we refrain from drawing a definitive classification for V148 at this stage. A comprehensive analysis incorporating these new data will be carried out in a future study.
We therefore consider the number of FUor-type spectra identified among the eight observed candidates to be $k=6$--7, where $k=7$ applies if V148 is ultimately confirmed as a FUor.

Assuming a parent sample of $N=18$ FUor candidates, among which $K$ objects truly exhibit FUor-type spectra, the probability of identifying $k$ FUors in a randomly selected subsample of $n=8$ objects follows a hypergeometric distribution,
\begin{equation}
P_{A|B}(n,k,N,K)=\frac{C_K^k C_{N-K}^{n-k}}{C_N^n},
\end{equation}
where event $B$ denotes the existence of $K$ FUor-type sources in the full sample and event $A$ represents the identification of $k$ FUor spectra in the observed subsample. Assuming a uniform prior probability for $K$, the posterior distribution is given by
\begin{equation}
p_{B|A}(n,k,N,K)=\frac{p_{A|B}(n,k,N,K)\,p_B(K,N)}{p_A(n,k,N)},
\end{equation}
with
\begin{equation}
p_A(n,k,N)=\sum_{K=k}^{N} p_{A|B}(n,k,N,K)\,p_B(K,N).
\end{equation}
For $N=18$, $n=8$, and $k=6$--7, we infer an expected value of $K/N \simeq 0.72-0.83$. The posterior probability that more than 12 objects in the parent sample exhibit FUor-type spectra reaches $63\%-90\%$, supporting the conclusion that infrared variability is an efficient way to identify FUor systems.

\section{Conclusions} \label{sec:conclusions}

In this work, we have presented near-infrared spectroscopic follow-up observations of four FUor candidates identified via infrared variability by \citet{2024MNRAS.532.2683L}, using the IGRINS instrument on Gemini South. Our main conclusions are as follows:

\begin{enumerate}
    \item Spectroscopic confirmation of FUor characteristics: All four sources (V10, V191, V450, and V978) exhibit classical FUor spectral signatures, including strong CO $\Delta v = 2$ absorption, pronounced H$_2$O absorption shaping the H-band continuum, Brackett line absorption, and weak metallic absorption features (Na I, Ca I), with little to no emission lines. Equivalent width diagnostics further support their FUor classification.

    \item Physical parameters from spectral modeling: Using a combined viscous and passively heated disk model fitted to the near-infrared spectra and SPHEREx photometry, we derived key properties of the sources. The inferred accretion rates ($\dot M \sim 10^{-5}$--$10^{-4}\,M_\odot\,\mathrm{yr^{-1}}$, assuming $M_* \sim 0.5\,M_\odot$) and high extinctions ($A_V \gtrsim 10$) are consistent with heavily embedded FUor systems, emphasizing the power of infrared surveys to uncover obscured outbursts missed by optical searches.  

    \item Validation of the fitting methodology: Application of the same modeling procedure to FU Orionis demonstrates that our approach reliably constrains fundamental parameters such as $\log_{10}(M\dot M)$ and $\sin i \sqrt{M_*}$ from near-infrared spectra alone, while parameters sensitive to optical or mid-infrared coverage (e.g., $R_\mathrm{in}$, $A_V$, $R_\mathrm{out}$, $\alpha$) may retain systematic uncertainties.

    \item Efficiency of infrared variability selection: Among the eight FUor candidates with available near-infrared spectra, six to seven exhibit bona fide FUor-type spectra, implying a high success rate ($72-83\%$) for selecting FUor outbursts based on infrared light curves. This highlights the effectiveness of mid-infrared variability surveys in identifying episodic accretion events, particularly for sources inaccessible to optical monitoring.
\end{enumerate}

Overall, our study confirms that infrared time-domain selection, combined with high-resolution spectroscopic follow-up and detailed disk modeling, provides a robust framework for identifying and characterizing FUor outbursts in young stellar objects. Future systematic studies of infrared-selected FUor populations, with broader wavelength coverage, will particularly benefit from optical spectra to better constrain extinction and inner disk properties, as well as mid-infrared spectra to model the passive disk.

\begin{acknowledgements}
This research is supported by NSFC funding (NSFC-12393814). 
Based on observations obtained at the international Gemini Observatory, a program of NSF NOIRLab (Program ID: GS-2024A-Q-113), which is managed by the Association of Universities for Research in Astronomy (AURA) under a cooperative agreement with the U.S. National Science Foundation on behalf of the Gemini Observatory partnership: the U.S. National Science Foundation (United States), National Research Council (Canada), Agencia Nacional de Investigaci\'{o}n y Desarrollo (Chile), Ministerio de Ciencia, Tecnolog\'{i}a e Innovaci\'{o}n (Argentina), Minist\'{e}rio da Ci\^{e}ncia, Tecnologia, Inova\c{c}\~{o}es e Comunica\c{c}\~{o}es (Brazil), and Korea Astronomy and Space Science Institute (Republic of Korea). 
This work used The Immersion Grating Infrared Spectrometer (IGRINS) was developed under a collaboration between the University of Texas at Austin and the Korea Astronomy and Space Science Institute (KASI) with the financial support of the US National Science Foundation under grants AST-1229522, AST-1702267 and AST-1908892, McDonald Observatory of the University of Texas at Austin, the Korean GMT Project of KASI, the Mt. Cuba Astronomical Foundation and Gemini Observatory.
The RRISA is maintained by the IGRINS Team with support from McDonald Observatory of the University of Texas at Austin and the US National Science Foundation under grant AST-1908892.
This publication also makes use of data products from the Spectro-Photometer for the History of the Universe, Epoch of Reionization and Ices Explorer (SPHEREx), which is a joint project of the Jet Propulsion Laboratory and the California Institute of Technology, and is funded by the National Aeronautics and Space Administration, and from NEOWISE,a project of the Jet Propulsion Laboratory/California Institute of Technology, funded by the Planetary Science Division of the National Aeronautics and Space Administration. In addition, this work makes use of data products from the Two Micron All Sky Survey (2MASS), a joint project of the University of Massachusetts and the Infrared Processing and Analysis Center/California Institute of Technology, funded by the National Aeronautics and Space Administration and the National Science Foundation. We also acknowledge the use of the SIMBAD database, operated at CDS, Strasbourg, France, and near-infrared photometry data provided by the VISTA VVV survey. 

\end{acknowledgements}

%

\bibliographystyle{aa} 
\bibliography{reference}


\begin{appendix}




\section{Fitting procedure} \label{sec:Fitting procedure}

The fitting procedure is based on Bayesian inference. The observational data used in the fitting consist of two components: 
(i) the ground-based near-infrared spectra and 
(ii) the SPHEREx low-resolution spectroscopic measurements.
Both datasets are jointly used to constrain the model parameters.

For a model with parameter $\vec{\theta_j}$, the likelihood of the combined dataset is defined as

\begin{equation}
p(\mathrm{data}|\vec{\theta}_j)
=
p(\mathrm{data}_{\mathrm{spec}}|\vec{\theta}_j)
\,
p(\mathrm{data}_{\mathrm{SPHEREx}}|\vec{\theta}_j),
\end{equation}

where $\mathrm{data}_{\mathrm{spec}}$ denotes the ground-based near-infrared spectra and 
$\mathrm{data}_{\mathrm{SPHEREx}}$ denotes the SPHEREx spectroscopic measurements.

Assuming independent Gaussian uncertainties, each likelihood term can be written as
\begin{equation}
p(\mathrm{data}_k|\vec{\theta}_j)
=
\frac{1}{\sqrt{(2\pi)^{n_k}}\prod_{i=1}^{n_k}\sigma_{k,i}}
\exp\left[
-\frac{1}{2}
\sum_{i=1}^{n_k}
\left(
\frac{F^{k}_{\mathrm{model},i}(\vec{\theta}_j)-F^{k}_{i}}
{\sigma_{k,i}}
\right)^2
\right],
\end{equation}
where the subscript $k$ refers to either the ground-based spectra or the SPHEREx data,
$F^{k}_{\mathrm{model},i}$ is the model flux averaged over the corresponding wavelength bin,
$F^{k}_{i}$ is the observed flux, and $\sigma_{k,i}$ is the associated uncertainty.

The total likelihood can therefore be expressed as
\begin{equation}
p(\mathrm{data}|\vec{\theta}_j)
\propto
\exp\left(-\frac{1}{2}\chi^2_{\mathrm{tot}}\right),
\end{equation}
with
\begin{equation}
\chi^2_{\mathrm{tot}}
=
\chi^2_{\mathrm{spec}}
+
\chi^2_{\mathrm{SPHEREx}}.
\end{equation}

The corresponding log-likelihood function is then written as
\begin{equation}
\ln \mathcal{L}(\vec{\theta})
=
-\frac{1}{2}\chi^2_{\mathrm{tot}}
+
\mathrm{const},
\end{equation}
where the additive constant is irrelevant for parameter inference and is therefore omitted in the MCMC sampling with 32 chains (walkers). The posterior distributions of the model parameters are explored using a MCMC sampler. For each source, the chains are evolved for 20,000 steps, and the first 5,000 steps are discarded as burn-in. The convergence of the chains is assessed using the integrated autocorrelation time ($\tau$), which is found to be in the range of about $10$–$20$ for all parameters across all sources.
For visualization, we use the corner.corner function from the Python corner package to generate the posterior distribution plots. A mild smoothing is applied with the parameters smooth = 1 and smooth1d = 1, corresponding to Gaussian smoothing of the two-dimensional and one-dimensional distributions, respectively. This smoothing is applied solely to suppress high-frequency sampling noise due to the finite length of the MCMC chains. The relatively small smoothing scale ensures that key features of the posterior distributions, such as peak locations and widths, are preserved.
\FloatBarrier

\section{Distribution, corner diagrams, and spectral Fits}
\label{sec:Distribution, corner diagrams, and spectral Fits}
We present the parameter distributions, corner diagrams, and near-infrared spectral fits for all sources. 
The layout and notation follow those described in Figs.~\ref{fig:V10_far_corner} and \ref{fig:V10_oms}.

\begin{figure}[h!]
    \centering
    \includegraphics[width=1.0\linewidth]{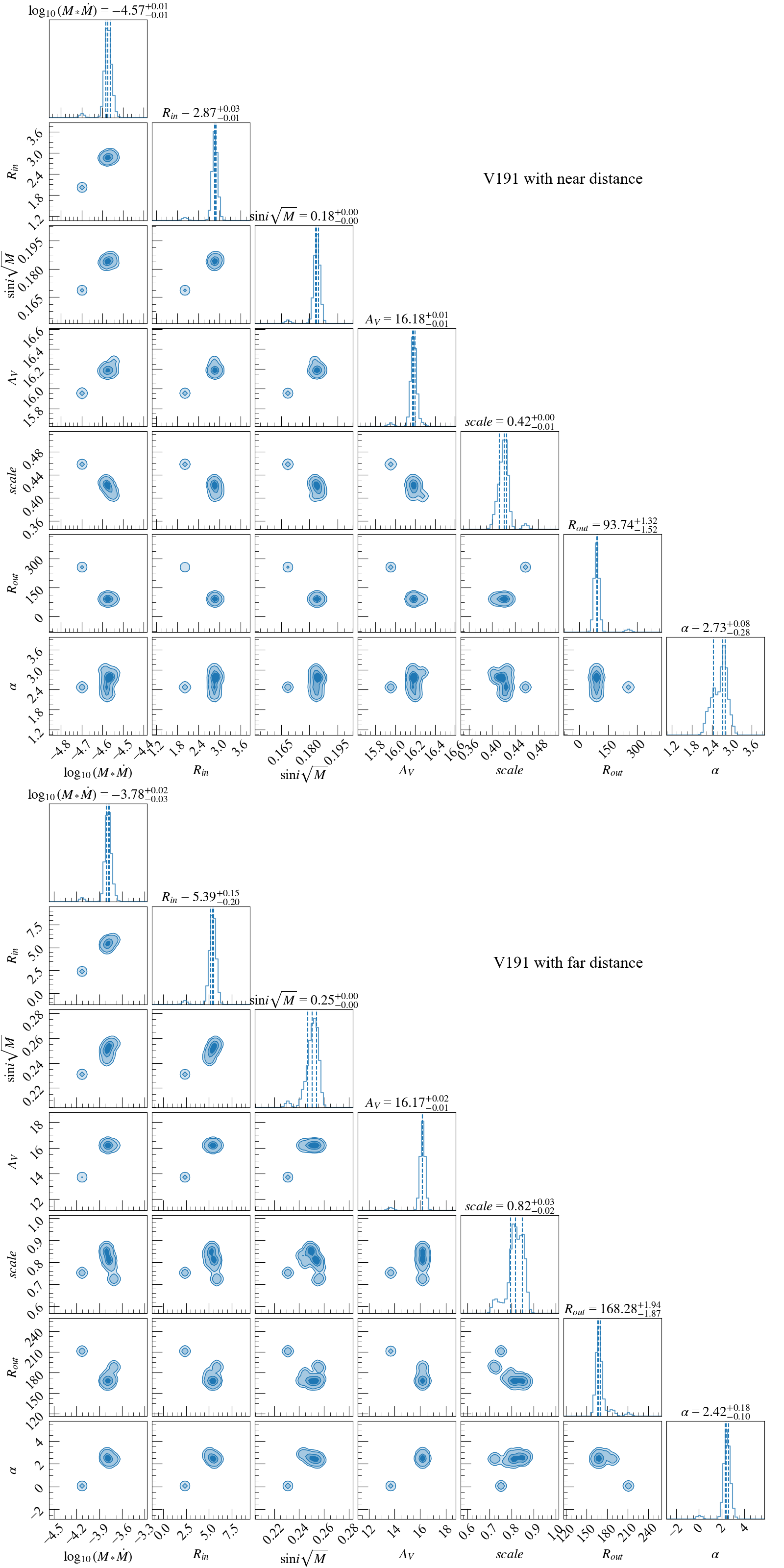}
    \caption{Distribution of parameters and corner diagram for V191. The upper panel shows the near solution, and the lower panel shows the far solution. The markings in the figure are consistent with Fig.~\ref{fig:V10_far_corner}.}
    \label{fig:corner_A1}
\end{figure}

\begin{figure}[h!]
    \centering
    \includegraphics[width=1.0\linewidth]{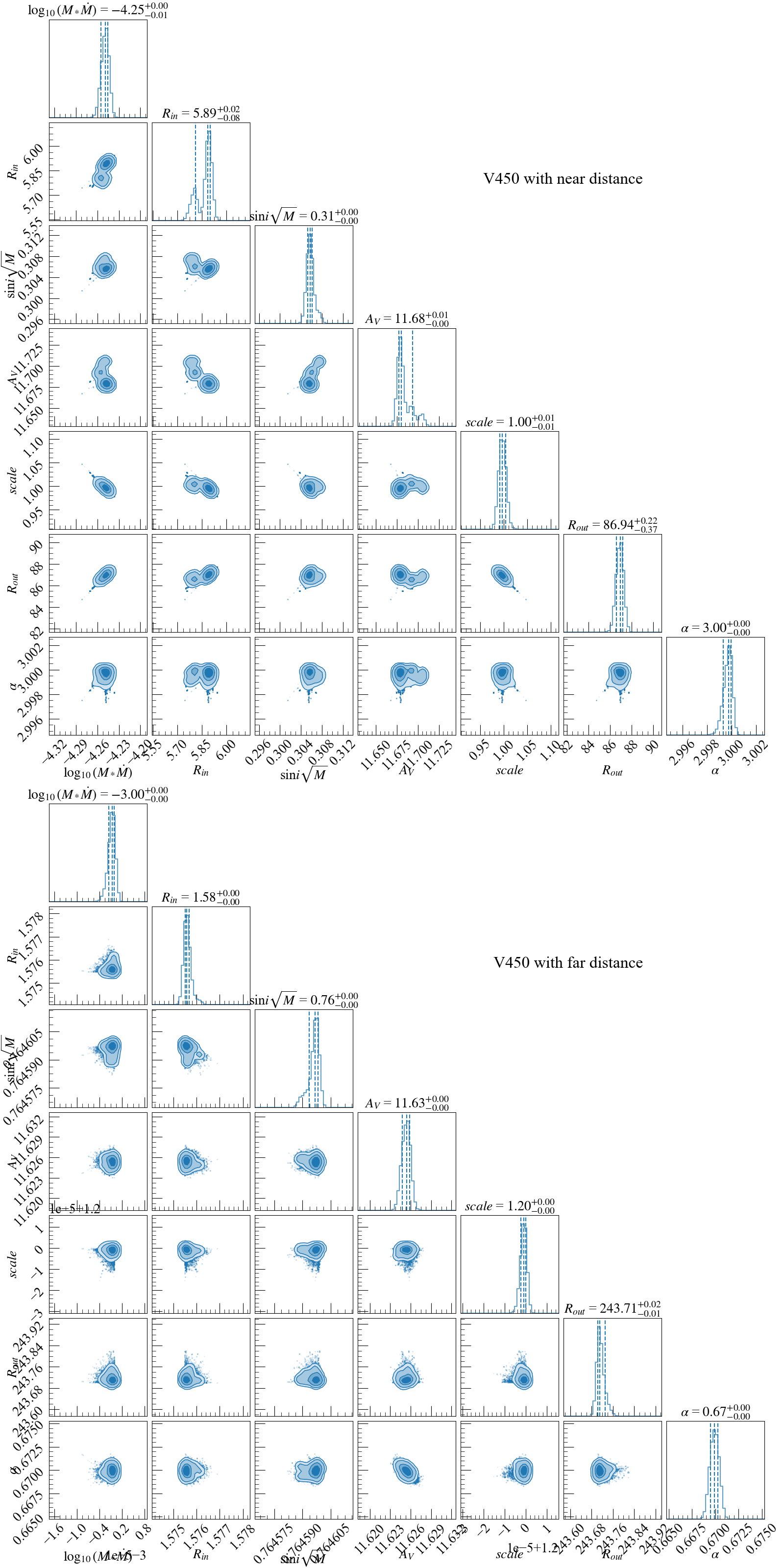}
    \caption{Distribution of parameters and corner diagram for V450. The upper panel shows the near solution, and the lower panel shows the far solution. The markings in the figure are consistent with Fig.~\ref{fig:V10_far_corner}.}
    \label{fig:corner_A2}
\end{figure}

\begin{figure}[h!]
    \centering
    \includegraphics[width=1.0\linewidth]{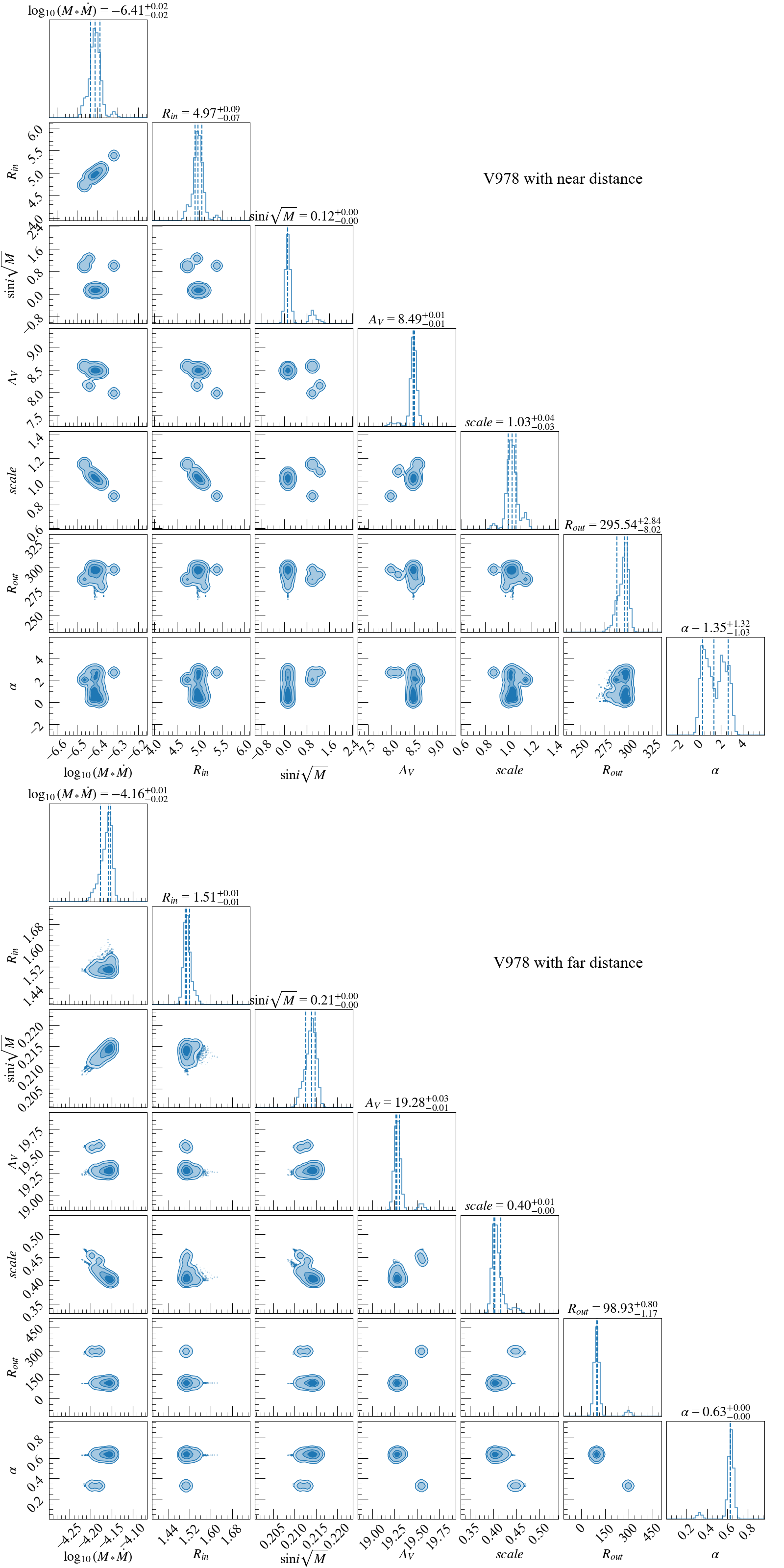}
    \caption{Distribution of parameters and corner diagram for V978. The upper panel shows the near solution, and the lower panel shows the far solution. The markings in the figure are consistent with Fig.~\ref{fig:V10_far_corner}.}
    \label{fig:corner_A3}
\end{figure}

\begin{figure}[h!]
    \centering
    \includegraphics[width=1.0\linewidth]{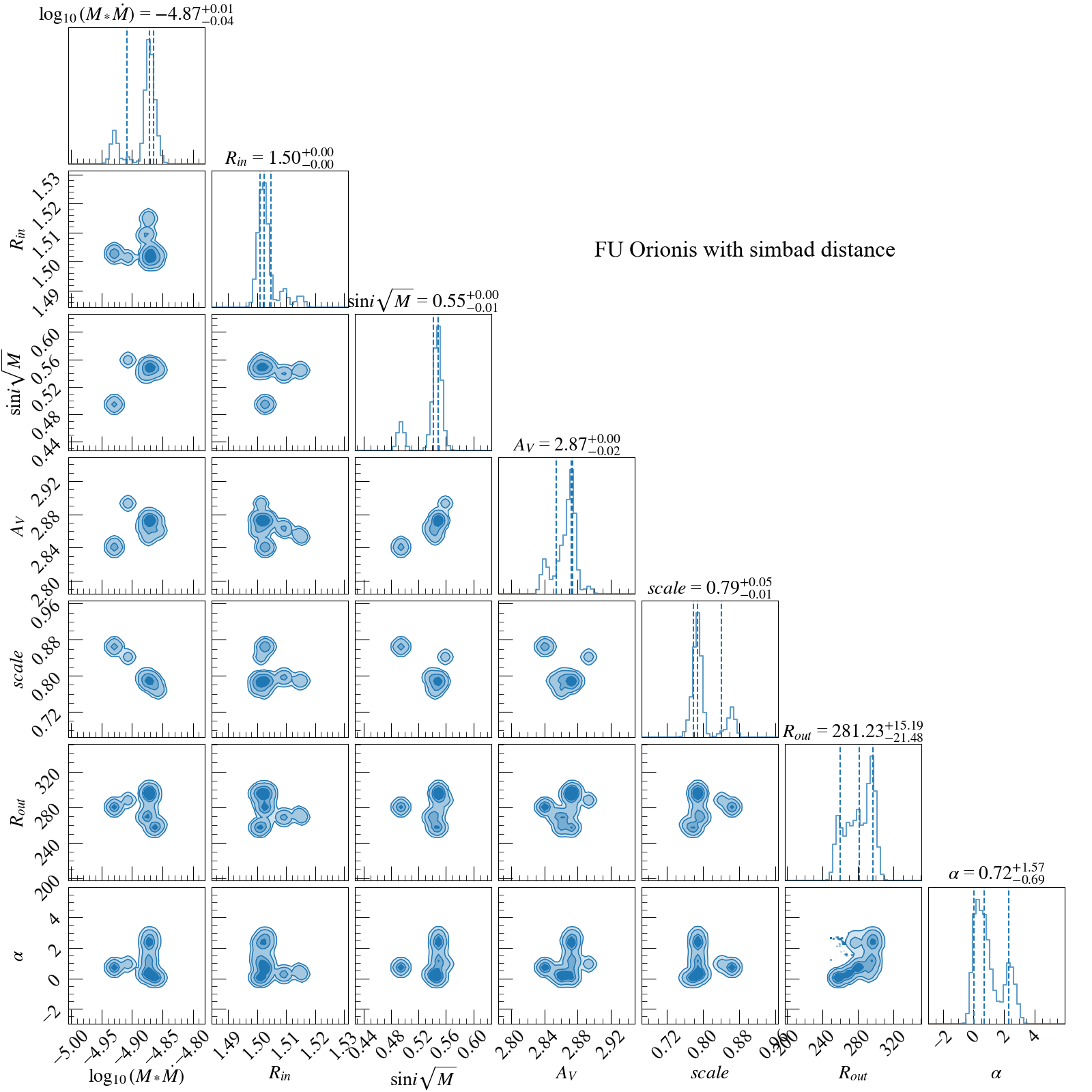}
    \caption{Distribution of parameters and corner diagram for FU Orionis. The upper panel shows the near solution, and the lower panel shows the far solution. The markings in the figure are consistent with Fig.~\ref{fig:V10_far_corner}.}
    \label{fig:corner_A4}
\end{figure}

\begin{figure*}
    \centering
    \includegraphics[width=1.0\linewidth]{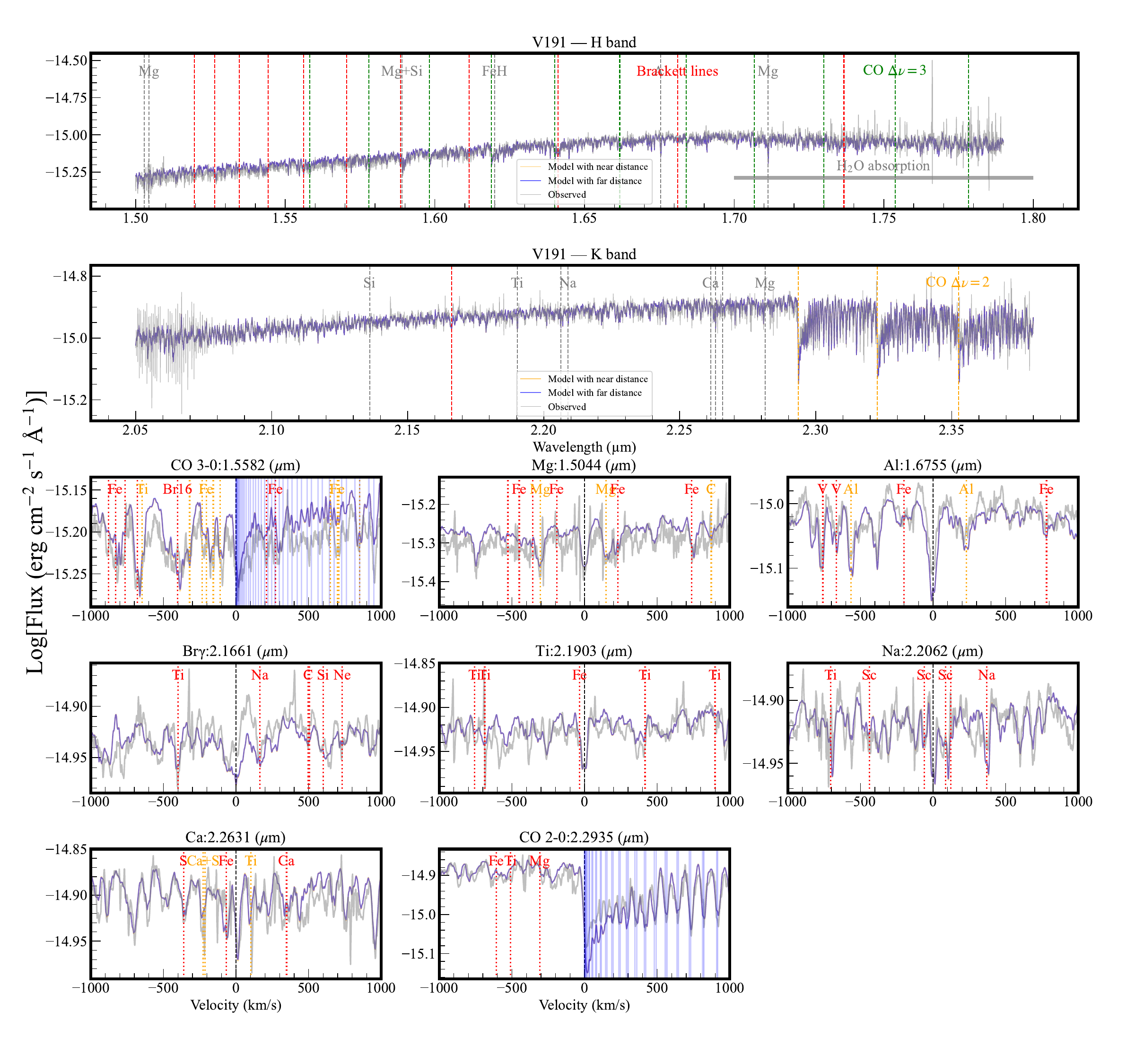}
    \caption{Near-infrared H- and K-band spectra of V191 compared with best-fitting disk models. 
    Panels are arranged as in Fig.~\ref{fig:V10_oms}.}
    \label{fig:V191_oms}
\end{figure*}

\begin{figure*}
    \centering
    \includegraphics[width=1.0\linewidth]{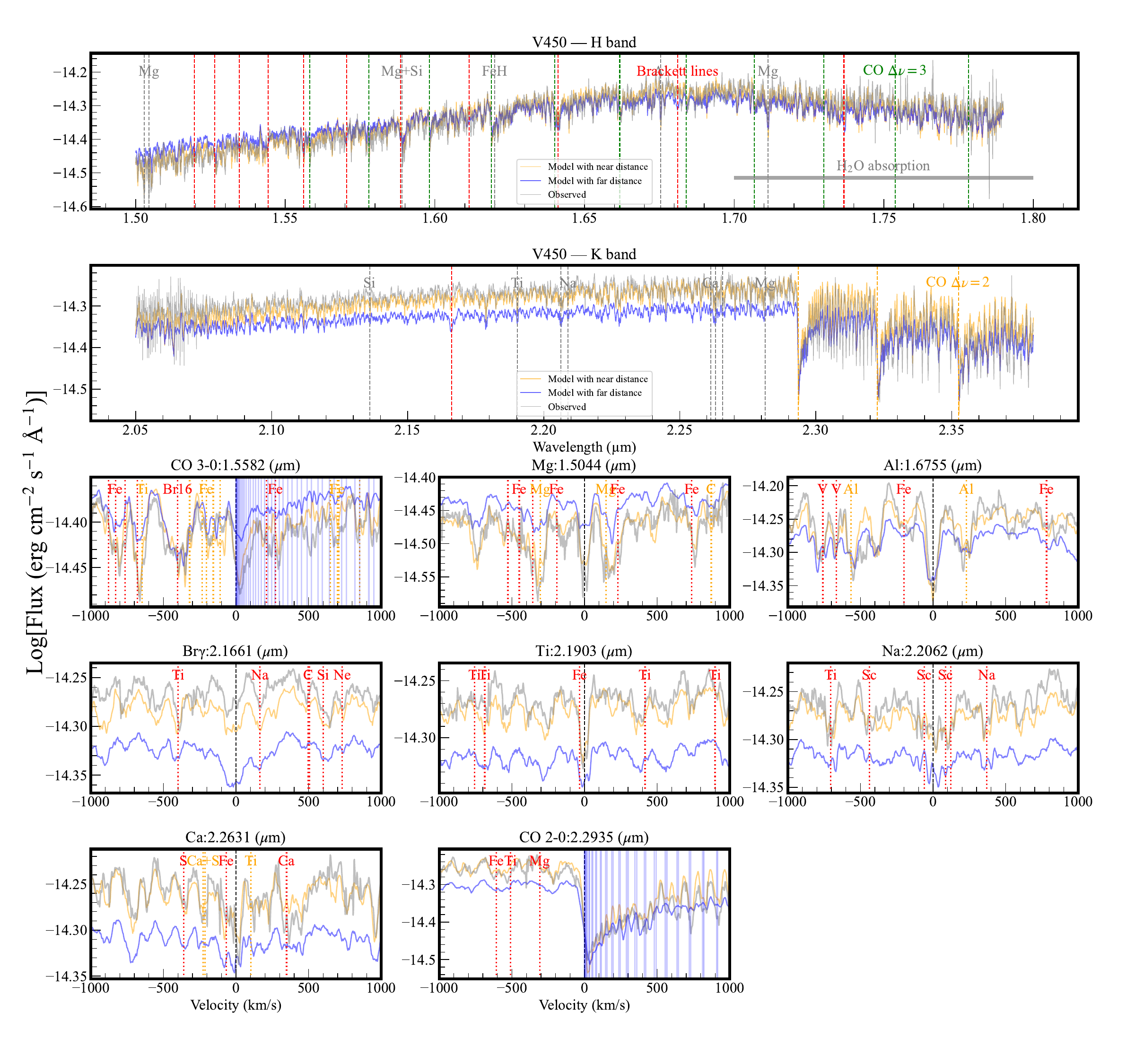}
    \caption{Near-infrared H- and K-band spectra of V450 compared with best-fitting disk models. 
    Panels are arranged as in Fig.~\ref{fig:V10_oms}.}
    \label{fig:V450_oms}
\end{figure*}

\begin{figure*}
    \centering
    \includegraphics[width=1.0\linewidth]{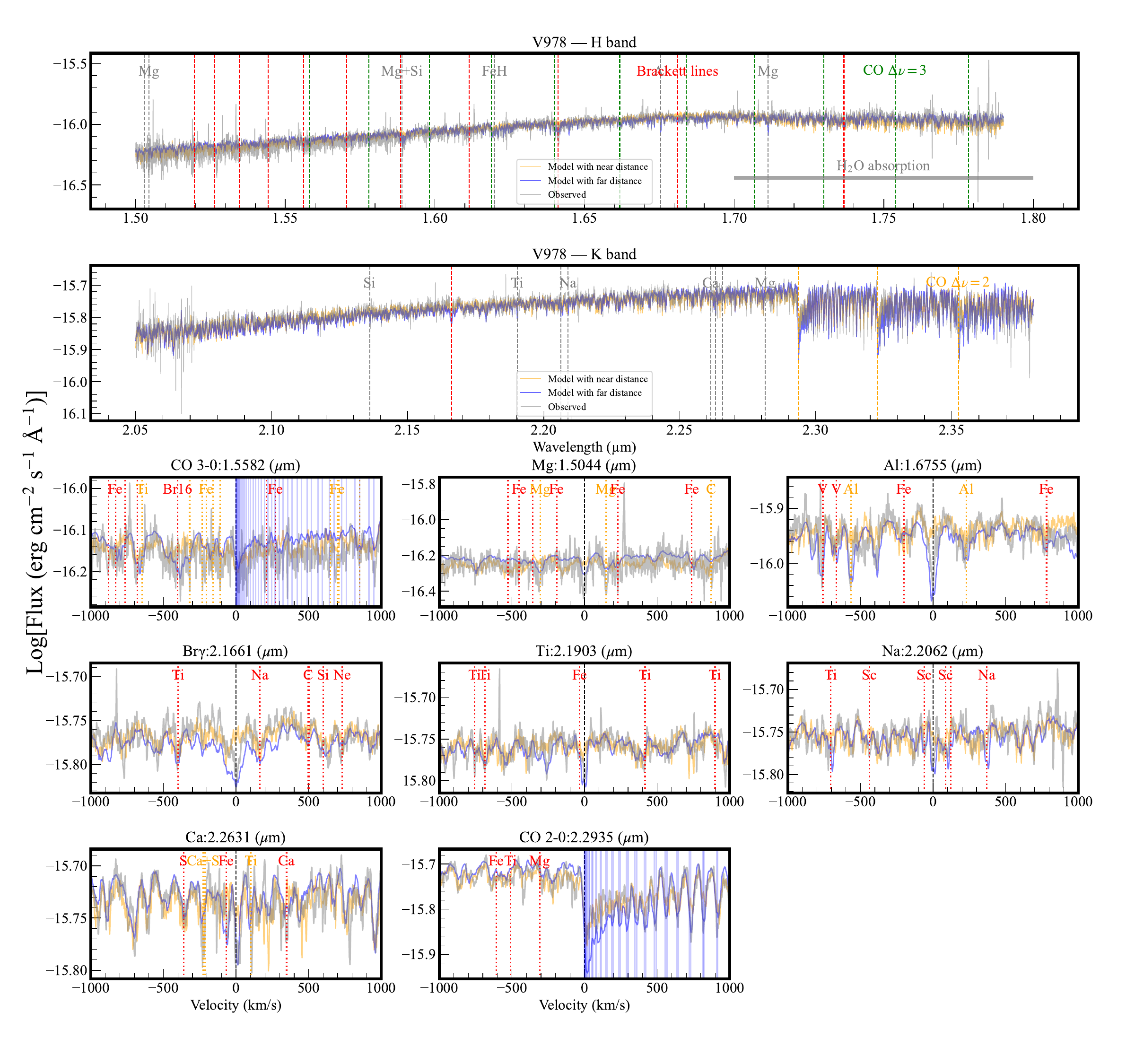}
    \caption{Near-infrared H- and K-band spectra of V978 compared with best-fitting disk models. 
    Panels are arranged as in Fig.~\ref{fig:V10_oms}.}
    \label{fig:V978_oms}
\end{figure*}

\begin{figure*}
    \centering
    \includegraphics[width=1.0\linewidth]{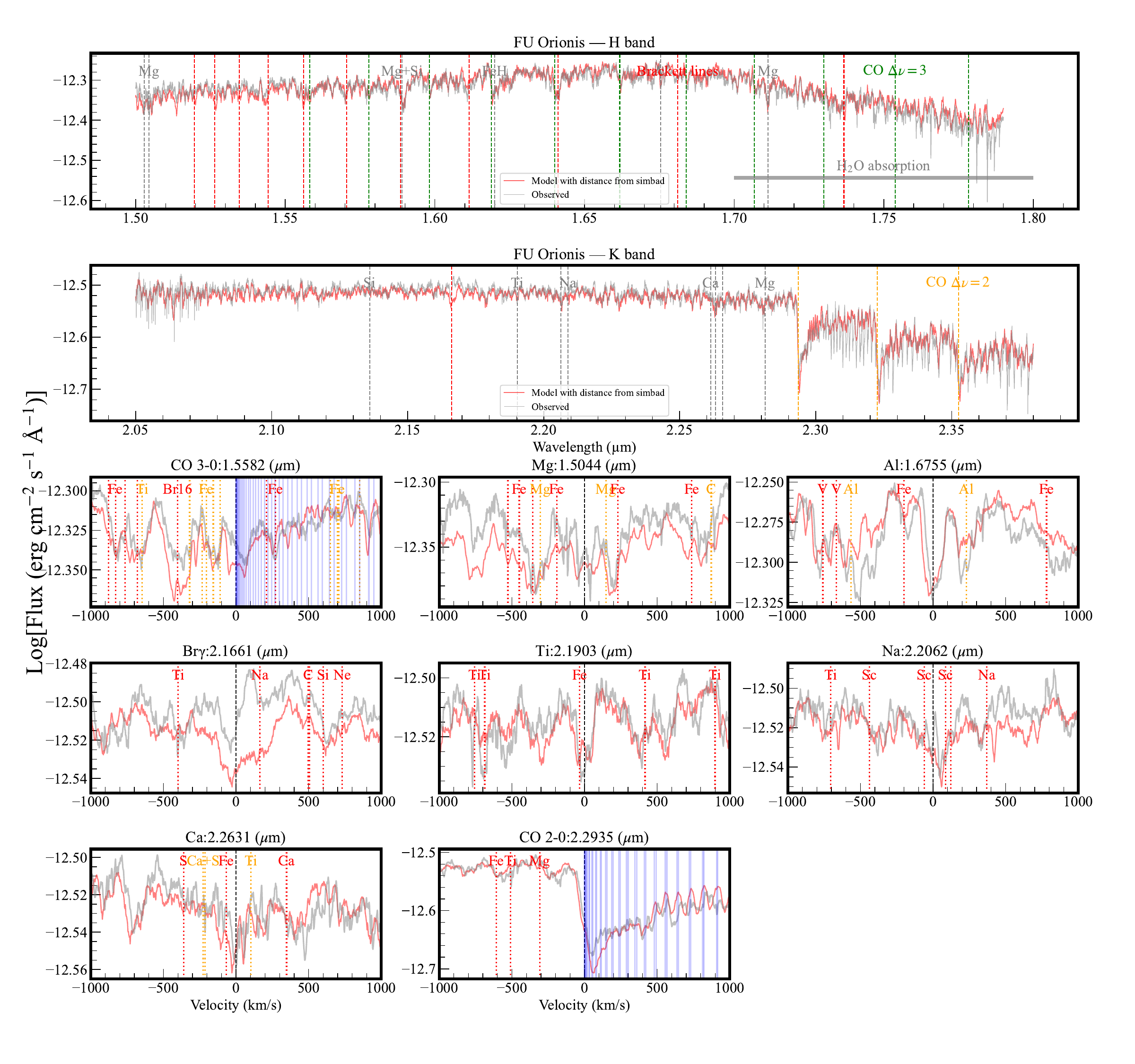}
    \caption{Near-infrared H- and K-band spectra of FU Orionis compared with best-fitting disk models. 
    Panels are arranged as in Fig.~\ref{fig:V10_oms}.}
    \label{fig:FU Orionis_oms}
\end{figure*}

\FloatBarrier

\end{appendix}

\end{document}